%% file: draft.tex
\documentclass[aps,prx,noeprint,twocolumn,showpacs,amsmath,amssymb]{revtex4-2}
\usepackage{amsmath}
\usepackage{graphicx}
\usepackage{subfigure}
\usepackage{epstopdf}
\usepackage{color}
\usepackage{multirow}
\usepackage{setspace}
\usepackage{overpic}
\usepackage{amssymb}
\usepackage{siunitx}
\usepackage{amsmath}
\usepackage[bookmarksnumbered, pdfstartview=FitH,colorlinks,urlcolor=blue, citecolor=blue,linkcolor=blue]{hyperref}
\usepackage{lineno}
\usepackage{bm}
\usepackage{rotating}
\usepackage{float}
\usepackage{booktabs}
\usepackage[utf8]{inputenc}
\let\oldequation\equation
\let\oldendequation\endequation

\renewenvironment{equation}
  {\linenomathNonumbers\oldequation}
  {\oldendequation\endlinenomath}
  
\begin{document}

\title{\bf \boldmath
Production and Decay Dynamics of the Charmed Baryon $\Lambda_c^+$\\ in $e^+e^-$ Annihilations near Threshold
}

\author{BESIII Collaboration}
\thanks{Full author list given at the end of the paper}

\vspace{4cm}

\date{\it \small \bf \today}

\begin{abstract}
The study of the charmed baryons is crucial for investigating the strong and weak interactions in the standard model and for gaining insights into the internal structure of baryons. 
In an $e^+e^-$ experiment the lightest charmed baryon, $\Lambda_c^+$, can be produced in pairs through the single-photon annihilation process. 
This process can be described by two complex electromagnetic form factors. The presence of a nonzero relative phase between these form factors gives rise to a transverse polarization of the charmed baryon and provides additional constraints on the dynamic parameters in the decays. 
In this article, we present the first observation of the transverse polarization of $\Lambda_{c}^{+}$ in the reaction $e^+e^- \to \Lambda_c^{+}\bar{\Lambda}_c^-$, based on $6.4~\text{fb}^{-1}$ of $e^{+}e^{-}$ annihilation data collected at center-of-mass energies between $4600$~MeV and $4951$~MeV with the BESIII detector.
The decay asymmetry parameters in the decays $\Lambda_c^+ \to pK_S^0$, $\Lambda\pi^+$, $\Sigma^0\pi^+$, and $\Sigma^+\pi^0$ are simultaneously extracted from the joint angular distributions. 
By independently extracting the above parameters for the process and its $C\!P$-conjugate counterpart, we derive both the strong and weak phase differences between different partial waves, and test several $C\!P$ observables.
The obtained results fully demonstrate the feasibility of exploiting transverse polarization at electron-positron colliders to search for $C\!P$-violating effects, opening up a new avenue for exploring $C\!P$ violation in the charmed baryon sector.
\end{abstract}

\newcommand{\BESIIIorcid}[1]{\href{https://orcid.org/#1}{\hspace*{0.1em}\raisebox{-0.45ex}{\includegraphics[width=1em]{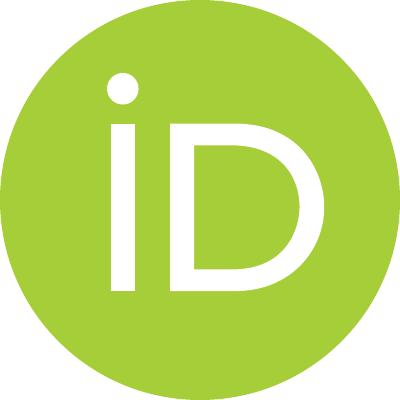}}}}

\maketitle

\oddsidemargin  -0.2cm
\evensidemargin -0.2cm

\section{Introduction}\label{sec:int}

Baryons, with three valence quarks, constitute the bulk of the visible matter around us. However, they are much less well understood than mesons, with only two valence quarks.
The increased complexity due to the additional quark makes precise theoretical predictions for baryons much more challenging, and many of the peculiar features of the strong interaction manifest themselves in the properties of the baryons. The rapid development of modern facilities, combining high precision with high energies, allows us to expand our horizon from protons and neutrons to baryons containing heavier strange, charm or beauty quarks. Of particular interest is the lightest charmed baryon, $\Lambda_c^+$, discovered in 1979~\cite{Abrams:1979iu}. Properties like charge, spin and parity are the same as for the proton, but instead of the valence quark composition $uud$, it is a strongly bound $udc$ system where the charm ($c$) quark is more than 250 times heavier than the light quarks. This feature should drastically impact its inner dynamics: In the large mass limit, i.e., when the quark mass $m_Q$ is substantially larger than the confinement scale $\Lambda_{QCD}$, the heavy quark acts like a nonrecoiling source of color and decouples from the nonperturbative dynamics of the light quarks. However, it is not completely established whether the charm quark can be considered heavy enough for this decoupling to apply. Investigations of charmed baryon structure and decay dynamics are therefore essential for exploring the strong and weak interactions within the standard model~(SM) of particle physics.

An approach to gain insights into hadron properties, which has been exploited successfully for more than 70 years, is that of electromagnetic form factors~(EMFFs)~\cite{Schonning:2023mge,Huang:2021xte,Pacetti:2014jai}. Spacelike form factors, studied in elastic electron-hadron scattering, have been rigorously studied for protons and, to some extent, neutrons, which has resulted in profound insights regarding structure and size~\cite{Punjabi:2015bba,Pacetti:2014jai,Atac:2021wqj}. 
How does the presence of a strange quark that is over 20 times heavier, or a charm quark that is over 250 times heavier, influence the structure of baryons? 
This problem is challenging since strange and charmed baryons have such short lifetimes ($\tau\approx10^{-10}$ s for strange and $\tau\approx10^{-13}$ s for charmed) that they are practically unfeasible as beams or targets in elastic scattering experiments. 
However, the rapid development of high-intensity and high-quality electron-positron collider experiments paves the way for measurements of EMFFs in the $e^+e^-$ annihilation process~\cite{Schonning:2023mge,Huang:2021xte,Pacetti:2014jai}. 
Recently, the BESIII experiment performed complete measurements of the EMFFs of strange $\Sigma^+$ \cite{BESIII:2023ynq} and $\Lambda$ \cite{BESIII:2025kyg} hyperons. 
Dispersive calculations based on Ref.~\cite{Mangoni:2021qmd} in the latter allowed for a determination of the rms charge radius. 
Charmed baryons are more challenging due to the large number of decay channels, each with low probability. 
To approach this intriguing problem requires dedicated methods, combining a number of charm baryon decay modes.

The formalism of the $e^+e^- \to \mathcal{B}\bar{\mathcal{B}}$ process is the same for strange and charmed spin-1/2 baryons.  The process is dominated by single-photon exchange, and the cross section can be described by the complex electric and magnetic form factors $G_E$ and $G_M$, which are functions of the four-momentum squared $q^2$ of the exchanged virtual photon. 
The joint angular distribution of the produced baryon and its subsequent decay is characterized by two parameters: the modulus of the ratio of the electric and the magnetic form factors $R = |G_E/G_M|$, and their relative phase $\Delta\Phi=\text{arg}(G_M)-\text{arg}(G_E)$~\cite{Chen:2019hqi}. Defining the angular parameter $\alpha_0$ in terms of the EMFF ratio $R$ as $\alpha_0=(|G_E|^2(1-v^2)-|G_M|^2)/(|G_E|^2(1-v^2)+|G_M|^2)$, where $v$ is the velocity of $\mathcal{B}$ in center-of-mass (c.m.) system, normalized to the speed of light, the transverse polarization can be parametrized as
\begin{equation}\label{hel:pt}
    \boldsymbol{\mathcal{P}_\mathcal{B}^y} (\cos\theta_{0})=\frac{\sqrt{1-\alpha^{2}_{0}}}{1+\alpha_0\cos^2\!\theta_0}\sin\theta_{0}\cos\theta_{0}\sin\Delta\Phi.
\end{equation}
Since the production proceeds predominantly via single-photon exchange, which conserves $C\!P$, $\Delta\Phi$ does not induce $C\!P$ asymmetry, but reflects the relative phase between $G_E$ and $G_M$.

A nonvanishing $\Delta\Phi$ gives rise to the transverse polarization of the baryon, which depends on the momentum direction of the baryon ($\theta_0$) in the $e^+e^-$ rest frame.
\begin{figure}[tp]
\centering
\includegraphics[width=0.9\linewidth]{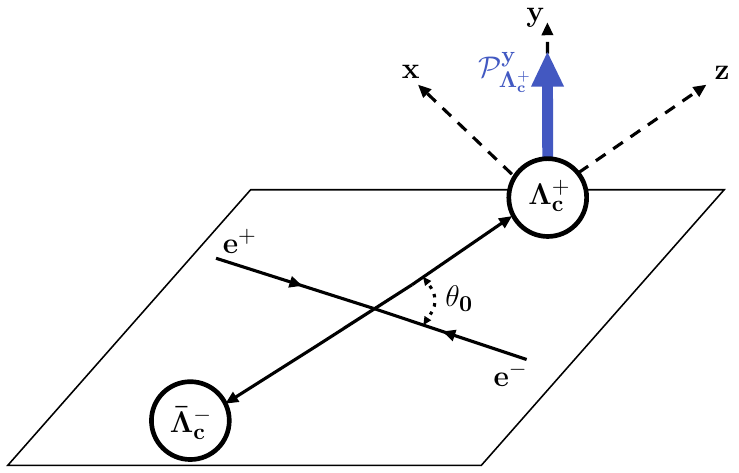}
	\caption{Definition of the coordinate system and kinematic variables for the process $e^+e^-\to\Lambda_c^+\bar{\Lambda}_c^-$. The production angle $\theta_0$ is measured between the positron beam and the $\Lambda_c^+$ momentum in the center-of-mass frame, and the polarization $\mathcal{P}_{\Lambda_c^+}^y$ is oriented perpendicular to the production plane.}
	\label{polarizationgeneration}
\end{figure}
Polarization in the production of hyperon-antihyperon pairs has been observed for $\Lambda$~\cite{BESIII:2018cnd,BESIII:2019nep,BESIII:2022qax, BESIII:2023euh,BESIII:2025kyg}, $\Sigma^{0,+}$~\cite{BESIII:2023cvk, BESIII:2023ynq, BESIII:2024nif} and $\Xi^{0,-}$~\cite{BESIII:2021ypr, BESIII:2022lsz, BESIII:2023drj} by the BESIII Collaboration. This finding has not only advanced our understanding of the hyperon structure, but it has also provided an instrument for precise measurements of hyperon decay parameters~\cite{Lee:1957qs} and a test of the symmetry under charge conjugation and parity transformation($C\!P$).
Similar to the pair production of hyperons, it is expected that the spin-1/2 charmed baryon $\Lambda_c^+$ via the $e^+e^- \to \Lambda_c^+\bar\Lambda_c^-$ production process can acquire polarization along the normal direction of the production plane, as illustrated in Fig.~\ref{polarizationgeneration}.
However, this effect has never been established in experiment.
A previous study of the transverse polarization of $\Lambda_c^+$ using data collected at a c.m. energy of $4600$~MeV at BESIII only had a statistical significance of 2.1 standard deviations~\cite{BESIII:2019odb}.

The amplitude for the weak decay $\Lambda_c^+ \to BP$, where $B$ and $P$ denote a spin-$1/2^+$ baryon and a pseudoscalar  meson, respectively, can be written as $\mathcal{M} = i \bar{u}_{f}(A-B\gamma_{5})u_{i}$. Here, $A$ and $B$ are  amplitudes to $S$ and $P$  angular momentum final states, arising from the parity-violating and parity-conserving components of the weak transition, respectively. The $u_{i}$ and $\bar{u}_{f}$ are spinors describing the initial and final baryons. The partial decay width related to the branching fraction is 
\begin{equation}\label{eqgroup}
\begin{split}
\Gamma_{BP}&=\frac{\hslash}{\tau_{\Lambda_{c}^{+}}}\mathcal{B}(\Lambda_{c}^{+}\to BP)\\
&=\frac{|\vec{p}_{c}|}{8\pi}\frac{(m_{\Lambda_{c}^{+}}+m_{B})^{2}-m_{P}^{2}}{m_{\Lambda_{c}^{+}}^{2}}(|A|^{2}+\kappa^2|B|^{2}),\\
\end{split}
\end{equation}
where $\kappa=|\vec{p}_{c}|/(E_B+m_B)$ while $E_B$ and $\vec{p}_{c}$ are the energy and momentum of the baryon $B$ in the rest frame of $\Lambda_{c}^{+}$. The mean lifetime of $\Lambda_{c}^{+}$is denoted as $\tau_{\Lambda_{c}^{+}}$. There are two independent parameters that describe angular distribution and polarization in the decay:
\begin{align}
    \alpha_{BP}&=\frac{2\kappa\text{Re}(A^*B)}  {|A|^{2}+\kappa^{2}|B|^{2}}\ , \\
    \Delta_{BP}&={\arg}\!\left[(A+\kappa B)(A^* - \kappa B^*)\right] \ ,
\end{align}
where $\alpha_{BP}$ and $\Delta_{BP}$ are the decay asymmetry parameters of $\Lambda_c^+ \to BP$.
Other decay asymmetry parameters $\beta_{BP}$ and $\gamma_{BP}$ are given by
\begin{align}
\beta_{BP}&=\sqrt{1-\alpha^2_{BP}}\sin\Delta_{BP}\,  \\
\gamma_{BP}&=\sqrt{1-\alpha^2_{BP}}\cos\Delta_{BP}\ .
\end{align}
However, for physics discussions, a more relevant quantity is the relative phase between $A$ and $B$ amplitudes
\begin{equation}
    \Delta_S=\text{arg}(A^*B)\ 
\end{equation}
and the two phases are related as
\begin{equation}
    \sin\Delta_{BP}=\frac{\alpha_{BP}}{\sqrt{1-\alpha^2_{BP}}}\tan\Delta_S\ .
\end{equation}

\begin{figure}[tp]
\centering
\includegraphics[width=0.9\linewidth]{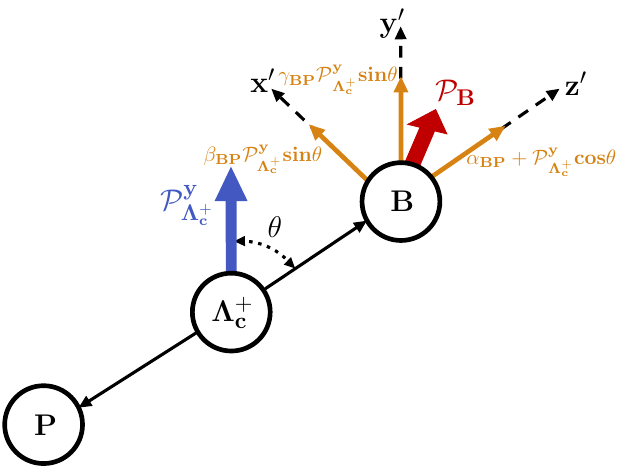}
	\caption{Generation and transfer of polarization in the weak decay $\Lambda_c^+ \to BP$~\cite{Wang:2024wrm}. The polarization of the daughter baryon $\boldsymbol{\mathcal{P}_{B}}$ depends on that of the parent $\boldsymbol{\mathcal{P}_{\Lambda_c^+}^y}$, the decay angle $\theta$, and the weak decay parameters $\alpha_{BP}$, $\beta_{BP}$, and $\gamma_{BP}$.}
	\label{polarization}
\end{figure}

The polarization of $\Lambda_c^+$, $\boldsymbol{\mathcal{P}_{\Lambda_c^+}^y}$, affects the kinematics of the subsequent decays and can be probed by the angular distributions of the secondary particles, as shown in Fig.~\ref{polarization}. 
The polarization of the daughter baryon, denoted as $\boldsymbol{\mathcal{P}_{B}}$, is expressed in the following formula
\begin{align}\small
    \frac{(\alpha_{BP} + \boldsymbol{\mathcal{P}_{\Lambda_c^+}^y}\cdot\boldsymbol{\hat{n}})\boldsymbol{\hat{n}} 
    + \beta_{BP}(\boldsymbol{\mathcal{P}_{\Lambda_c^+}^y}\times\boldsymbol{\hat{n}}) 
    + \gamma_{BP}\boldsymbol{\hat{n}}\times(\boldsymbol{\mathcal{P}_{\Lambda_c^+}^y}\times\boldsymbol{\hat{n}})}{1 + \alpha_{BP} \boldsymbol{\mathcal{P}_{\Lambda_c^+}^y}\cdot\boldsymbol{\hat{n}}}, 
\end{align}
where $\boldsymbol{\hat{n}}$ is the unit vector along the baryon flight direction in the rest frame of $\Lambda_c^+$. 

Various phenomenological models taking into account nonperturbative effects and nonfactorizable contributions in QCD have been developed to calculate the decay asymmetries of the charmed baryons. 
The development of these models can be roughly divided into three stages. 
First, the CLEO Collaboration measured $\Lambda_{c}^{+}\to\Lambda\pi^{+}$ in 1990~\cite{CLEO:1990unw}. 
Models such as the covariant confined quark model (CCQM)~\cite{Korner:1992wi,Ivanov:1997ra}, the pole model~\cite{Xu:1992vc,Cheng:1991sn,Cheng:1993gf,Zenczykowski:1993jm,Zenczykowski:1993hw}, and current algebra (CA)~\cite{Datta:1995mn,Sharma:1998rd} were applied to calculate decay asymmetry parameters. 
The decay asymmetry parameters $\alpha_{pK^{0}_{S}}$, $\alpha_{\Lambda\pi^{+}}$, $\alpha_{\Sigma^{0}\pi^{+}}$, and $\alpha_{\Sigma^{+}\pi^{0}}$ were calculated using different models and varied significantly, particularly with regard to the sign differences, as shown in Table~\ref{restable}. 
The beginning of the second stage is marked by the measurement of the BESIII Collaboration in 2019~\cite{BESIII:2019odb}, which was the first to report $\alpha_{pK^{0}_{S}}$ and $\alpha_{\Sigma^{0}\pi^{+}}$, as well as the improved results of $\alpha_{\Lambda\pi^{+}}$ and $\alpha_{\Sigma^{+}\pi^{0}}$.
As the central value of $\alpha_{pK^{0}_{S}}$ at BESIII is positive with a large uncertainty, the sign of $\alpha_{pK^{0}_{S}}$ remained a puzzle at that time, although most of the predictions quoted negative values.
However, the signs of the other three decay asymmetries were confirmed to be negative, which was further supported by the Belle Collaboration~\cite{Belle:2022uod,Belle:2022bsi}.
Subsequently, the BESIII Collaboration first reported the $\Lambda_{c}^{+}\to\Xi^{0}K^{+}$ decay asymmetry parameters in 2024~\cite{BESIII:2023wrw}, which denote a pure $W$-exchange process. 
Calculations are very difficult because of the nonperturbative nature and the nonfactorization problem in this decay channel. 
A persistent tension exists between the branching fraction and the decay asymmetry parameters, as theoretical calculations and experimental measurements do not align perfectly.
To resolve this tension, studies of the $\Lambda_{c}^{+}\to\Xi^{0}K^{+}$ decay bring a valuable new perspective centered on the $\Delta_S$
relative phase between $P$- and $S$-wave amplitudes, which plays a crucial role in tests of the $C\!P$ symmetry~\cite{Yang:2025orn}. 
In particular, the prerequisite for $C\!P$ tests using $\alpha_{BP}$ is $\tan\Delta_{S}\ne0$. 
The origin of the $\Delta_S$ phase is a strong interaction in the final state.
This final-state interaction could also affect the value of the decay asymmetry parameter $\alpha_{BP}$. Therefore some recent theoretical works~\cite{Geng:2023pkr,Zhong:2024zme,Zhong:2024qqs} used it to improve the description of the data.
The most notable change is seen for $\alpha_{pK^{0}_{S}}$, whose value is calculated to be close to 0~\cite{Zhong:2024qqs}. 
However, in 2024, the $\alpha_{pK^{0}_{S}}$ result from the LHCb Collaboration indicated that it is negative. 
The BESIII Collaboration can further improve the measurement of these decay asymmetries using a larger sample compared to the previous study, in particular to confirm the sign of $\alpha_{pK^{0}_{S}}$.

\begin{table*}[tp]
	\footnotesize
	\centering
	\caption{Theoretical calculations and experimental measurements of the decay asymmetry parameters. The superscript $a$ means the model only considered SU(3) symmetry, and $b$ means the model considered the contribution of both SU(3) symmetry and SU(3) broken symmetry.}
	\label{restable}
	\begin{tabular}{c c c c c}
		\hline\hline
		\raisebox{0.2ex}{Calculations or experiments} & \raisebox{0.75ex}{$\alpha_{pK^{0}_{S}}$} & \raisebox{0.75ex}{$\alpha_{\Lambda\pi^{+}}$} & \raisebox{0.75ex}{$\alpha_{\Sigma^{0}\pi^{+}}$} & \raisebox{0.75ex}{$\alpha_{\Sigma^{+}\pi^{0}}$}\\
		\hline
		K{\"o}rner(1992), CCQM~\cite{Korner:1992wi}             & $-0.10$                 & $-0.70$                          & $ 0.70$                      & $ 0.71$ \\
		Xu(1992), Pole~\cite{Xu:1992vc}                         & $ 0.51$                 & $-0.67$                          & $ 0.92$                      & $ 0.92$ \\
		Cheng, Tseng(1992), Pole~\cite{Cheng:1991sn}            & $-0.49$                 & $-0.96$                          & $ 0.83$                      & $ 0.83$ \\
		Cheng, Tseng(1993), Pole~\cite{Cheng:1993gf}            & $-0.49$                 & $-0.95$                          & $ 0.78$                      & $ 0.78$ \\
		{\.Z}enczykowski(1994), Pole~\cite{Zenczykowski:1993jm} & $-0.66$                 & $-0.99$                          & $ 0.39$                      & $ 0.39$ \\
		{\.Z}enczykowski(1994), Pole~\cite{Zenczykowski:1993hw} & $-0.90$                 & $-0.86$                          & $-0.76$                      & $-0.76$ \\
		Alakabha Datta(1995), CA~\cite{Datta:1995mn}            & $-0.91$                 & $-0.94$                          & $-0.47$                      & $-0.47$ \\
		Ivanov(1998), CCQM~\cite{Ivanov:1997ra}                 & $-0.97$                 & $-0.95$                          & $ 0.43$                      & $ 0.43$ \\
		Sharma(1999), CA~\cite{Sharma:1998rd}                   & $-0.99$                 & $-0.99$                          & $-0.31$                      & $-0.31$ \\
		Geng(2019), SU(3)~\cite{Geng:2019xbo}                   & $-0.89^{+0.26}_{-0.11}$ & $-0.87\pm0.10$                   & $-0.35\pm0.27$               & $-0.35\pm0.27$ \\
		Zou(2020), CA~\cite{Zou:2019kzq}                        & $-0.75$                 & $-0.93$                          & $-0.76$                      & $-0.76$ \\
        Zhong (2022), SU(3)$^{a}$~\cite{Zhong:2022exp}          & $-0.51\pm0.21$          & $-0.75\pm0.01$                   & $-0.47\pm0.03$               & $-0.47\pm0.03$ \\
		Zhong (2022), SU(3)$^{b}$~\cite{Zhong:2022exp}          & $-0.29\pm0.24$          & $-0.75\pm0.01$                   & $-0.47\pm0.03$               & $-0.47\pm0.03$ \\
        Liu(2023), Pole~\cite{Liu:2023dvg}                      & $-0.81\pm0.05$          & $-0.75\pm0.01$                   & $-0.47\pm0.01$               & $-0.45\pm0.04$ \\ 
        Liu(2023), LP~\cite{Liu:2023dvg}                        & $-0.68\pm0.01$          & $-0.75\pm0.01$                   & $-0.47\pm0.01$               & $-0.45\pm0.04$ \\
        Geng(2023), SU(3)~\cite{Geng:2023pkr}                   & $-0.40\pm0.49$          & $-0.75\pm0.01$                   & $-0.47\pm0.02$               & $-0.47\pm0.02$ \\
        Zhong(2024), TDA~\cite{Zhong:2024qqs}                   & $ 0.01\pm0.24$          & $-0.76\pm0.01$                   & $-0.48\pm0.02$               & $-0.48\pm0.02$ \\
        Zhong(2024), IRA~\cite{Zhong:2024qqs}                   & $ 0.03\pm0.24$          & $-0.76\pm0.01$                   & $-0.48\pm0.02$               & $-0.48\pm0.02$ \\
        Zhong(2024), TDA~\cite{Cheng:2024lsn}                   & $-0.74\pm0.03$          & $-0.76\pm0.01$                   & $-0.47\pm0.01$               & $-0.47\pm0.01$ \\
        Zhong(2024), IRA~\cite{Cheng:2024lsn}                   & $-0.74\pm0.03$          & $-0.76\pm0.01$                   & $-0.47\pm0.01$               & $-0.47\pm0.01$ \\
		\hline
		CLEO(1990)~\cite{CLEO:1990unw}                          & \raisebox{0.5ex}{...}   & $-1.0^{+0.4}_{-0.1}$             & \raisebox{0.5ex}{...}        & \raisebox{0.5ex}{...} \\
		ARGUS(1992)~\cite{ARGUS:1991yzs}                        & \raisebox{0.5ex}{...}   & $-0.96\pm0.42$                   & \raisebox{0.5ex}{...}        & \raisebox{0.5ex}{...} \\
		CLEO(1995)~\cite{CLEO:1995qyd}                          & \raisebox{0.5ex}{...}   & $-0.94^{+0.21+0.12}_{-0.06-0.06}$& \raisebox{0.5ex}{...}        & $-0.45\pm0.31\pm0.06$ \\
		FOCUS(2006)~\cite{FOCUS:2005vxq}                        & \raisebox{0.5ex}{...}   & $-0.78\pm0.16\pm0.19$            & \raisebox{0.5ex}{...}        & \raisebox{0.5ex}{...} \\
		BESIII(2019)~\cite{BESIII:2019odb}                      & $0.18\pm0.43\pm0.14$    & $-0.80\pm0.11\pm0.02$            & $-0.73\pm0.17\pm0.07$        & $-0.57\pm0.10\pm0.07$ \\
		BELLE(2022)~\cite{Belle:2022uod,Belle:2022bsi}          & \raisebox{0.5ex}{...}   & $-0.755\pm0.005\pm0.003$         & $-0.463\pm0.016\pm0.008$     & $-0.48\pm0.02\pm0.02$ \\
        LHCb(2024)~\cite{LHCb:2024tnq}                          & $-0.754\pm0.008\pm0.006$& $-0.785\pm0.006\pm0.003$         & \raisebox{0.5ex}{...}        & \raisebox{0.5ex}{...} \\
		PDG Fit(2025)~\cite{ParticleDataGroup:2024cfk}          & $-0.754\pm0.010$           & $-0.768\pm0.015$                 & $-0.466\pm0.018$             & $-0.484\pm0.027$ \\
		\hline\hline
	\end{tabular}
\end{table*}

Violation of the $C\!P$ symmetry lies at the heart of some of the most fundamental problems in physics. 
The presence of $C\!P$ violation has been established in meson decays~\cite{KTeV:1999kad,NA48:1999szy,LHCb:2019hro,BaBar:2004gyj,Belle:2004nch}.
In 2019, the LHCb Collaboration observed a difference in time-integrated $C\!P$ asymmetries between the decays $D^0\to K^+K^-$ and $D^0\to\pi^+\pi^-$~\cite{LHCb:2019hro}, providing the first observation of direct $C\!P$ violation in the charm sector.
This finding established that all three quark sectors$-$strange, bottom, and charm$-$exhibit $C\!P$-violating effects, albeit at very different magnitudes.
Recently, the LHCb Collaboration reported the first observation of direct $C\!P$ violation in a baryon decay~\cite{LHCb:2025ray}, $\Lambda_b^0\to pK^-\pi^+\pi^-$, marking a significant milestone in the study of $C\!P$ violation. 
Meanwhile, there are ongoing efforts to explore $C\!P$ violation in other baryon decays~\cite{LHCb:2016yco,LHCb:2024yzj,BESIII:2026mye}. 
Despite these advancements, the role of different partial waves and their interference behavior in the decay process in $C\!P$ violation remains an open question, and answering it is crucial for precise theoretical predictions and experimental searches.
In the SM, $C\!P$ violation in the baryon sector originates from the interference of amplitudes carrying different weak and strong phases, typically arising from contributions of tree and penguin diagrams. 
Under a $C\!P$ transformation, weak phases change sign while strong phases remain invariant. 
Consequently, a nonzero $C\!P$ asymmetry requires the simultaneous presence of a nonzero shift for both weak and strong phases between different diagrams.
Inspired by the strategy widely used in hyperon decays~\cite{BESIII:2023jhj}, $C\!P$ violation can be experimentally probed through observables constructed from the interference between the $S$- and $P$-wave amplitudes, defined as 
$A^{\alpha_{BP}}_{C\!P}=\frac{\alpha_{BP}+\bar{\alpha}_{BP}}{\alpha_{BP}-\bar{\alpha}_{BP}}$, $\text{tan}\phi_{C\!P}=\frac{\beta_{BP}+\bar{\beta}_{BP}}{\alpha_{BP}-\bar{\alpha}_{BP}}$ and $\text{tan}\Delta_{S}=\frac{\beta_{BP}-\bar{\beta}_{BP}}{\alpha_{BP}-\bar{\alpha}_{BP}}$, where $\bar\beta_{BP}=\sqrt{1-\bar\alpha_{BP}^2}\sin\bar\Delta_{BP}$.
The first two observables, $A^{\alpha_{BP}}_{C\!P}$ and $\tan\phi_{C\!P}$, are $C\!P$-violating observables, while $\tan\Delta_S$ characterizes the strong-phase difference between the amplitudes involved.

In the charmed baryon sector, Cabibbo-suppressed~(CS) decays, such as $\Lambda_c^+\to\Lambda K^+$, provide a promising channel to search for $C\!P$ violation, as the tree and penguin amplitudes contribute with distinct weak and strong phases in the SM. 
However, studies of such CS decays require an extremely large data sample, which makes the observation of $C\!P$ violation experimentally challenging~\cite{LHCb:2024tnq}.
Within the SM, the Cabibbo-favored~(CF) decays are dominated by tree-level amplitudes that share the same weak phase, with penguin contributions being highly suppressed. 
As a result, the $C\!P$-violating observables constructed from the decay parameters, such as $A_{C\!P}^{\alpha}$ and $\tan\phi_{C\!P}$, are expected to vanish. 
Thus, searching for nonzero effects of these observables provides a sensitive probe to study the $C\!P$-violation mechanisms beyond the SM.
On the other hand, the angular analysis framework, as proposed in this study for $C\!P$ violation and measuring both strong and weak phase differences, can be validated via CF decay channels, in which the derived transverse polarizations can be used to enhance the sensitivities of $C\!P$-violating observables. 
This framework provides a critical methodological foundation for subsequent $C\!P$-violation searches in CS decays.

In this paper, we observe for the first time the transverse polarization of the $\Lambda_{c}^{+}$ baryon in $e^{+}e^{-} \to \Lambda_c^+\bar\Lambda_c^-$ and obtain the sine of the phase difference~($\Delta\Phi$) between the EMFFs at 13 c.m. energies ranging from $4600$ to $4951$~MeV, with an overall significance exceeding 10 standard deviations. 
In the fit, the parameters $\alpha_{0}$ and $\sin\Delta\Phi$ are directly determined from the data.
At the decay level, the decay asymmetry parameters $\langle\alpha_{BP}\rangle$ and $\langle\Delta_{BP}\rangle$ are directly determined in three CF channels: $\Lambda_{c}^{+}\to\Lambda\pi^{+},~\Sigma^{0}\pi^{+}\text{,~and}~\Sigma^{+}\pi^{0}$. 
The $\langle\alpha_{BP}\rangle$ and $\langle\Delta_{BP}\rangle$ mean $C\!P$ average values, which are defined as $(\alpha_{BP}-\bar{\alpha}_{BP})/2$ and $(\Delta_{BP}-\bar{\Delta}_{BP})/2$.
Then $\langle\beta_{BP}\rangle$, $\langle\gamma_{BP}\rangle$, and phase differences between the $S$ and $P$ wave, $\langle\Delta_S\rangle$, can be derived from the measured decay asymmetry parameters $\langle\alpha_{BP}\rangle$ and $\langle\Delta_{BP}\rangle$. 
For the decay $\Lambda_{c}^{+}\to pK^{0}_{S}$, only the parameter $\langle\alpha_{BP}\rangle$ is extracted since the direction of the proton polarization vector is not measured.
The analysis is performed using a multidimensional angular study of the cascade decays of $\Lambda_c^+$, following the formalism described in Ref.~\cite{BESIII:2019odb}. In addition to the primary signal channels, the decay $\Lambda_c^+\to pK^-\pi^+$, which features a clean topology and a large branching fraction, is included to improve the sensitivity to the transverse polarization measurement. Unless otherwise specified, charge-conjugate channels are implied throughout this paper.

\section{Experimental apparatus and data sets}\label{sec:sample}
The Beijing Spectrometer III~(BESIII) detector~\cite{BESIII:2009fln} records symmetric $e^+e^-$ collisions provided by the Beijing Electron Positron Collider~(BEPCII) storage ring~\cite{Yu:2016cof} in the c.m. energy range from 1.84 to 4.95~GeV, with a peak luminosity of $1.1 \times 10^{33}\;\text{cm}^{-2}\text{s}^{-1}$ achieved at $\sqrt{s} = 3.773\;\text{GeV}$. 
BESIII has collected large data samples in this energy region~\cite{BESIII:2020nme,Lu2020-hf,Zhang2022-id}. 
The cylindrical core of the BESIII detector covers 93\% of the full solid angle and consists of a helium-based multilayer drift chamber~(MDC), a plastic scintillator time-of-flight system~(TOF), and a CsI(Tl) electromagnetic calorimeter~(EMC), which are all enclosed in a superconducting solenoidal magnet providing a 1.0~T magnetic field. The solenoid is supported by an octagonal flux-return yoke with resistive-plate-counter muon-identification modules interleaved with steel. 
The charged-particle momentum resolution at $1~{\rm GeV}/c$ is $0.5\%$, and the ${\rm d}E/{\rm d}x$ resolution is $6\%$ for electrons from Bhabha scattering. 
The EMC measures photon energies with a resolution of $2.5\%$ ($5\%$) at $1$~GeV in the barrel (end-cap) region. 
The time resolution in the TOF barrel region is 68~ps, while that in the end-cap region is 110~ps. The end-cap TOF system was upgraded in 2015 using multigap resistive-plate-chamber technology, providing a time resolution of 60~ps~\cite{Guo2017-of,Li2017-zt,Cao2020-vl}. 
Data analyzed in this work were collected at 13 energy points above the $\Lambda^+_c\Bar{\Lambda}^-_c$ threshold. 
The integrated luminosity of the datasets analyzed in this work is estimated to be $6.4~\mathrm{fb}^{-1}$~\cite{BESIII:2022dxl,BESIII:2022ulv}.

Simulated data samples produced with a {\sc geant4}-based~\cite{GEANT4:2002zbu} Monte Carlo (MC) package, which includes the geometric description of the BESIII detector and the detector response, are used to determine detection efficiencies and to estimate backgrounds. 
The simulation models the beam energy spread and initial-state radiation (ISR) in the $e^+e^-$ annihilations with the generator {\sc kkmc}~\cite{Jadach:2000ir,Jadach:1999vf}. 
The inclusive simulation includes the production of open charm processes, the ISR production of vector charmonium(like) states, and the continuum processes incorporated in {\sc kkmc}~\cite{Jadach:2000ir,Jadach:1999vf}. 
All particle decays are modeled with {\sc evtgen}~\cite{Lange:2001uf,Ping:2008zz} using branching fractions either taken from the Particle Data Group (PDG)~\cite{ParticleDataGroup:2024cfk}, when available, or otherwise estimated with {\sc lundcharm}~\cite{Chen:2000tv,Yang:2014vra}. 
Final-state radiation~(FSR) from charged final-state particles is incorporated using the {\sc photos} package~\cite{Barberio:1990ms}. 
The phase-space signal simulation is generated
according to a pure phase-space model, assuming a uniform distribution of the available kinematic phase space without any dynamical amplitudes or polarization effects for the process $e^{+}e^{-}\to\Lambda_c^{+}\bar{\Lambda}_c^{-}$, followed by the decay of $\Lambda_{c}^{+}$ into the signal final states and the inclusive decay of $\bar{\Lambda}_{c}^{-}$, which accounts for all possible decay modes.
The signal simulation is estimated using the process model described in the helicity formalism with the decay asymmetry parameters measured in this work or cited from the other measurements~\cite{BESIII:2022qax,ParticleDataGroup:2024cfk}.

\section{Measurement Method}\label{sec:method}
\subsection{Event selection and \texorpdfstring{$\Lambda_c^+$}{Lambdac} signal\\ in the beam-constrained mass spectrum}\label{sec:evt}

A single tag approach~\cite{BESIII:2017kqg} is employed, in which
an event is accepted as an exclusive $\Lambda_c^+\bar{\Lambda}_c^-$ final state when either a $\Lambda_c^+$ or its antiparticle is reconstructed in one of the decay channels studied, using all particles on the same side as the reconstructed baryon.
Four two-body hadronic decays$-$$\Lambda_c^+ \to pK_S^0$, $\Lambda \pi^+$, $\Sigma^0\pi^+$, and $\Sigma^+\pi^0$$-$and one three-body decay, $\Lambda_c^+ \to pK^-\pi^+$, are studied following Refs.~\cite{BESIII:2017kqg,BESIII:2022xne}. 
The intermediate states $K_S^0$, $\Lambda$, $\Sigma^0$, and $\Sigma^+$ are reconstructed by combining $\pi^+\pi^-$, $p\pi^-$, $\gamma\Lambda$, and $p\pi^0$ candidates, respectively.

Charged tracks detected in the MDC are required to satisfy $|\cos\theta| < 0.93$, where $\theta$ is the polar angle defined relative to the symmetry axis of the MDC (the $z$ axis). 
Additionally, the tracks that do not come from a $K^0_S$ meson or $\Lambda$ baryon must have a distance of closest approach to the interaction point (IP) satisfying $|r_z| < 10~\text{cm} and \quad |r_{xy}| < 1~\text{cm}$, where $r_z$ and $r_{xy}$ are the distances along the beam axis and in the transverse plane, respectively.

Particle identification~(PID) for charged tracks combines measurements of the energy deposited in the MDC~(d$E$/d$x$) and the flight time in the TOF to form likelihoods $\mathcal{L}(h)~(h=p, K,\pi)$ for each hadron $h$ hypothesis.
Tracks are identified as protons if their likelihoods satisfy $\mathcal{L}(p)>\mathcal{L}(K)$ and $\mathcal{L}(p)>\mathcal{L}(\pi)$. Charged kaons and pions are identified by comparing the likelihoods of the hypotheses
$\mathcal{L}(K)>\mathcal{L}(\pi)$ and $\mathcal{L}(\pi)>\mathcal{L}(K)$, respectively.
Each track is used only once in the reconstruction of $\Lambda_c^+$ decays, so potential overlaps between protons and kaons or pion identification are avoided.

Neutral showers are reconstructed in the EMC. Showers not associated with any charged track are identified as photon candidates. 
The deposited energy of each shower in the EMC must satisfy the following conditions: $E_{\text{barrel}}>25$ MeV for $|\cos\theta|<0.80$ and $E_{\text{end-cap}}>50$ MeV for $0.86<|\cos\theta|<0.92$. 
The EMC time difference from the event start time is required to be less than 700 ns to suppress electronic noise and showers unrelated to the event, where the event start time corresponds to the time at which the $e^+e^-$ collision occurs.
As verified with simulated signal samples, this requirement retains the vast majority of genuine signal showers and introduces only a negligible loss of signal efficiency.
The $\pi^0$ candidates are reconstructed from photon pairs with invariant masses satisfying: $115~\text{MeV}/c^2<M(\gamma\gamma)<150~\text{MeV}/c^2$.

Candidates for $K^0_S$ and $\Lambda$ decays are reconstructed from $\pi^+\pi^-$ and $p\pi^-$ combinations, respectively. 
For these tracks, the distances of closest approach to the IP must satisfy: $|r_z|<20~\text{cm}$. 
There is no distance constraint required in the transverse plane. The charged pions in the $K^0_S$ reconstruction are not subject to the PID requirements described earlier, while the proton in the $\Lambda$ reconstruction is required to satisfy the PID criteria. 
The two final-state tracks are constrained to originate from a common decay vertex by requiring the vertex fit to satisfy: $\chi^2<100$. Furthermore, the decay vertex must be separated from the IP by a distance of at least twice the fitted vertex resolution. 
To select $K^0_S$ and $\Lambda$ candidates, the invariant masses of the combinations must satisfy $487~\text{MeV}/c^2<M(\pi^+\pi^-)<511~\text{MeV}/c^2$ and $1111~\text{MeV}/c^2<M(p\pi^-)<1121~\text{MeV}/c^2$. 
These mass windows correspond to approximately 3 times the standard deviation on either side of the known masses.

To suppress the peaking backgrounds for some decay channels, like $\Lambda_{c}^{+}\to \Lambda l^{+}\nu_{l}(l=e/\mu)$ background in the $\Lambda_{c}^{+}\to\Lambda\pi^{+}$ channel, and $\Lambda_{c}^{+}\to \Lambda\pi^{+}$ background in the $\Lambda_{c}^{+}\to\Sigma^{0}\pi^{+}$ channel, additional requirements are applied. 
Positron PID uses the measured information in the MDC, TOF, and EMC. The combined likelihoods ($\mathcal{L}'$) under the positron, pion, and kaon hypotheses are obtained.
For the $\Lambda_{c}^{+}\to\Lambda\pi^{+}$ channel, it is required that $\mathcal{L}'(e)<0.05$ and $\mathcal{L}(\pi)>0.001$ are satisfied to suppress the $\Lambda_{c}^{+}\to\Lambda e^{+}\nu_{e}$ background, which rejects roughly 95\% of these backgrounds with only a negligible signal loss.
For the $\Lambda_{c}^{+}\to\Lambda\mu^{+}\nu_{\mu}$ background in the $\Lambda_{c}^{+}\to\Sigma^{0}\pi^{+}$ channel, the information from the muon counter detector is utilized, and the penetrated depth of the $\pi^{+}$ candidates is required to be less than $5~\text{cm}$.
These selections effectively reduce peaking backgrounds $\Lambda_{c}^{+}\to\Lambda\mu^{+}\nu_{\mu}$ by 80\% while keeping over 80\% of the signal events.
A kinematic variable, the energy difference $\Delta E\equiv E_{\Lambda_{c}^{+}}-E_{\text{beam}}$, is used to suppress background, where $E_{\Lambda_{c}^{+}}$ is the total measured energy of the $\Lambda_{c}^{+}$ candidate, and $E_{\text{beam}}$ is the beam energy.
For the $\Lambda^+_c\to pK^-\pi^+$, $pK^0_S$, $\Lambda\pi^+$, and $\Sigma^+\pi^0$ decays the requirements are $\Delta E\in(-29,26)~\text{MeV}$, $\Delta E\in(-21,18)~\text{MeV}$, $\Delta E\in(-23,21)~\text{MeV}$, and $\Delta E\in(-67,32)~\text{MeV}$, respectively.
For the $\Lambda_{c}^{+}\to\Sigma^{0}\pi^{+}$ channel, an additional photon can be selected from the process $\Lambda_{c}^{+}\to\Lambda\pi^{+}$, and its $\Delta E$ is expected to be larger than zero. 
The requirement $\Delta E\in(-33,20)~\text{MeV}$ removes about 54\% of this type of backgrounds while retaining more than 98\% of signal events.

A large background for the channel $\Lambda_c^+ \to \Sigma^+\pi^0$ is from $\Lambda_c^+ \to \Lambda \pi^+$ combined with a random photon.  
This background, which may pose hidden dangers to the angular analysis, necessitates the application of a four-constraint kinematic fit to the $\Lambda_{c}^{+}\to\Sigma^+\pi^{0}$, where $\Sigma^+$ decays to proton and $\pi^{0}$, $\pi^{0}\to\gamma\gamma$. 
The quality of the fit, $\chi^2$, is calculated by constraining two $\pi^{0}$s, $\Sigma^+$, and $\Lambda_{c}^{+}$ to their world average masses quoted from PDG~\cite{ParticleDataGroup:2024cfk}. Maximizing $N^2_S/(N_S+N_B)^{3/2}$~\cite{LHCb:2021chn,LHCb:2022ogu} leads to the requirement $\chi^2<12$, where $N_S$ and $N_B$ represent the signal and background yields, respectively.

\begin{figure}[htbp]
	\centering
	\includegraphics[width=\linewidth]{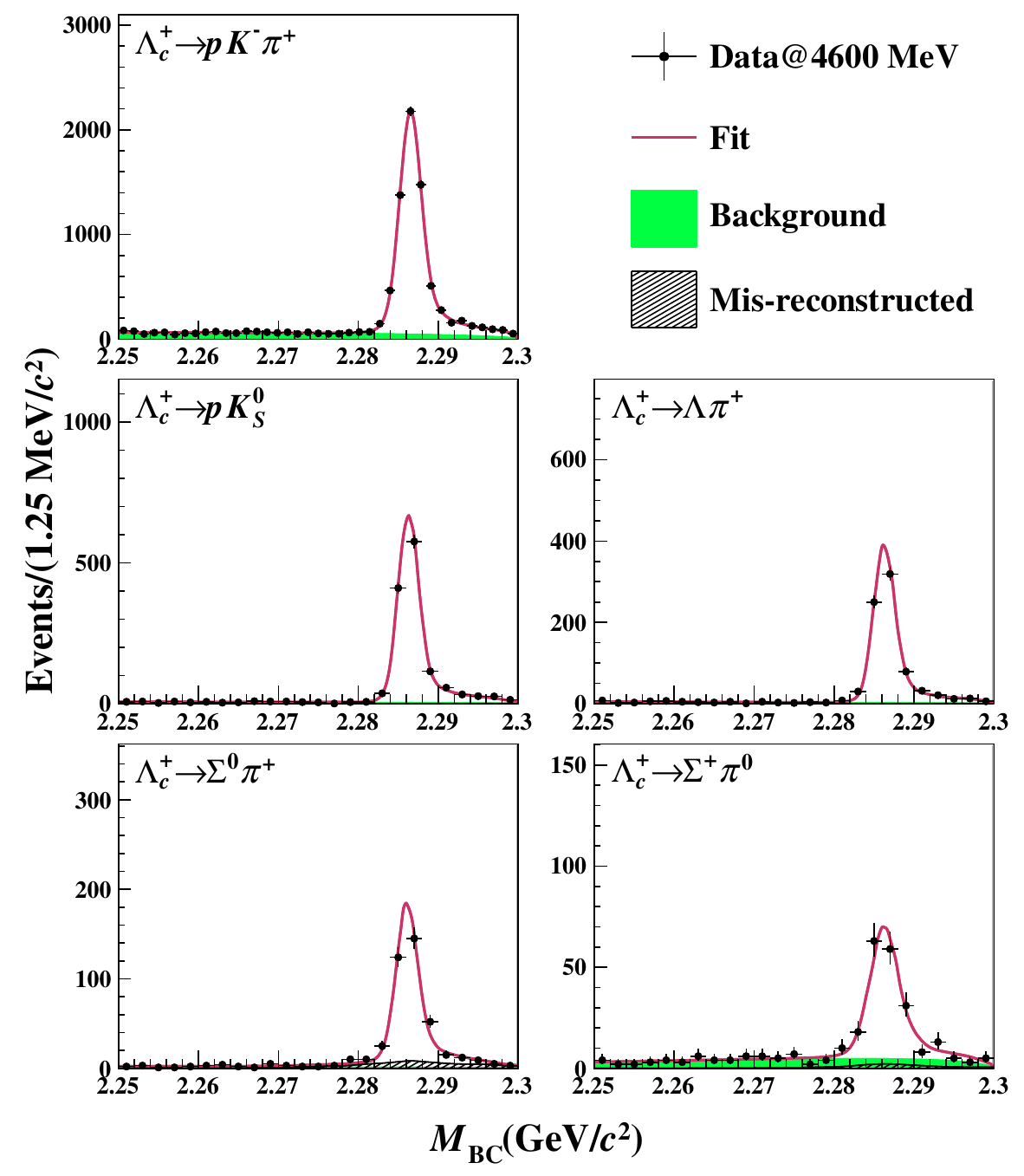}
	\caption{Beam-constrained mass spectrum for the $\Lambda_c^+$ signal at $4600$~MeV. Each panel shows a different signal channel as labeled. The black points with error bars are data. The black shaded region indicates the mis-reconstructed signal events. The green region represents the fitted combinatorial background.} \label{fig:combine6}
\end{figure}

After applying the above selection criteria, the beam-constrained mass $M_{\text{BC}}\equiv\sqrt{E_{\text{beam}}^{2}/c^{4}-|\vec{p}_{\Lambda_{c}^{+}}|^{2}/c^{2}}$, is defined to identify signal candidates, and $\vec{p}_{\Lambda_{c}^{+}}$ is the momentum of the $\Lambda_{c}^{+}$ candidates calculated in the $e^{+}e^{-}$ rest frame. 
An unbinned maximum-likelihood fit is performed on the distribution of $M_{\text{BC}}$. 
The shapes of both correctly and mis-reconstructed signal events are modeled with the simulation, convolved with a Gaussian function to account for the resolution difference between the data and the simulation. 
The fitted Gaussian parameters indicate only a small additional smearing with respect to the simulation.
The matching algorithm that relates generator level and reconstructed information classifies events as correctly reconstructed if the angle between the reconstructed and true photon momentum directions from the simulation is below 10 degrees; otherwise, they are mis-reconstructed. 
The background shapes are described by an ARGUS function~\cite{ARGUS:1990hfq}, except in the case of the $\Lambda_c^+ \to pK^-\pi^+$ channel at c.m. energies larger than $4740$~MeV, where a second-order polynomial function is employed to model the background.
The fit results are shown in Fig.~\ref{fig:combine6} for all channels at $4600$~MeV.
The obtained signal yields, combining all energy points, are $50083\pm233$, $9588\pm96$, $5734\pm74$, $2486\pm48$ and $1224\pm37$ in the $\Lambda_{c}^{+}\to pK^{-}\pi^{+}$,~$pK^{0}_{S},~\Lambda\pi^{+},~\Sigma^{0}\pi^{+}\text{,~and}~\Sigma^{+}\pi^{0}$ channels, respectively, within the signal region of $M_{\text{BC}}\in[2.282,2.291]~\text{GeV}/c^{2}$.

\subsection{Helicity system and angular distribution definitions}
\begin{figure}[htbp]
	\centering
	\includegraphics[width=\linewidth]{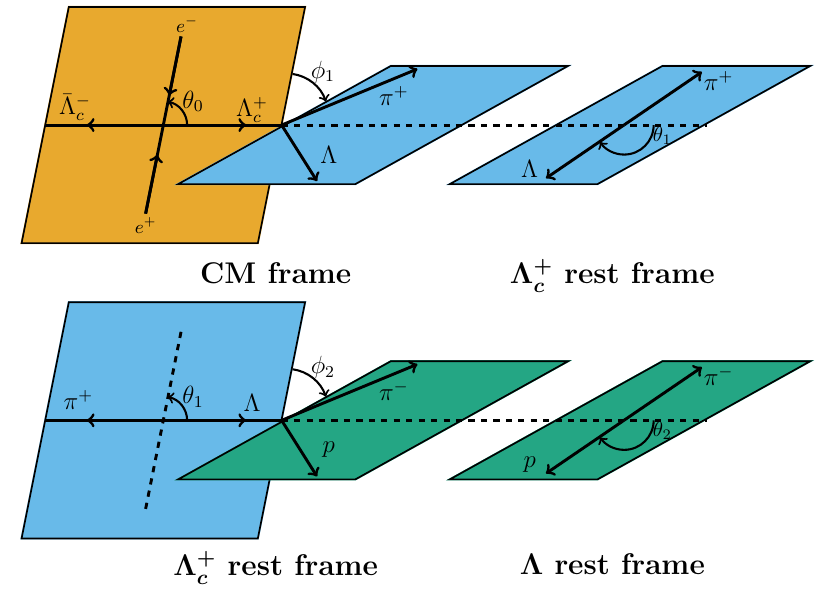}
	\caption{Definitions of the helicity frames and related angles for $e^{+}e^{-}\to\Lambda_{c}^{+}\bar{\Lambda}_c^{-}, \Lambda_{c}^{+}\to\Lambda\pi^{+}$, and $\Lambda\to p\pi^{-}$.}
	\label{fig:angledef}
\end{figure}

A helicity system and helicity angles are employed in constructing the amplitude for $e^{+}e^{-}\to\Lambda_c^+\bar{\Lambda}_c^-$, $\Lambda_c^+\to BP$. 
Taking the channel $\Lambda_c^+\to\Lambda\pi^{+}$ as an example,  a kinematic diagram of this decay is shown in Fig.~\ref{fig:angledef}. 
For the process $e^{+}e^{-}\to\Lambda_c^+\bar{\Lambda}_c^-$, the momenta of the $\Lambda_c^+$ and $\bar{\Lambda}_c^-$ baryons are defined in the c.m. system of the $e^{+}e^{-}$ pair, and $\theta_{0}$ is the polar angle of the $\Lambda_c^+$ with respect to the $e^+$ beam direction in the c.m. system. 
For the helicity system describing $\Lambda_c^+\to\Lambda\pi^{+}$ decay, the momenta of the $\Lambda$ and the $\pi^{+}$ are defined in the $\Lambda_c^+$ frame. 
The angle $\phi_{1}$ is that between the $e^{+}\Lambda_c^+$ and $\Lambda\pi^{+}$ planes, and $\theta_{1}$ is the polar angle of the $\Lambda$ momentum defined in the rest frame of the $\Lambda_c^+$ baryon with respect to the direction of the $\Lambda_c^+$ in the c.m. frame.
For the helicity system describing the $\Lambda\to p\pi^-$, $\phi_{2}$ and $\theta_{2}$ adopt similar definitions. 

The joint angular distribution of the single-side decay chains can be derived from polarization transmission~\cite{Chen:2019hqi} or, equivalently, the helicity density matrix~\cite{Perotti:2018wxm}. 
The angular distribution of the decay chain $e^+e^-\to \gamma^*\to \Bar{\Lambda}_c^-\Lambda_c^+(\to \Lambda(\to p\pi^-)\pi^+)$ is 
\begin{equation}
	\label{hel:formulalmdpi}
    \small
	\begin{split}
	&\frac{\text{d}\varGamma}{\text{d}\cos\theta_{0}~\text{d}\cos\theta_{1}~\text{d}\cos\theta_{2}~\text{d}\phi_{1}~\text{d}\phi_{2}}\\
	&\propto1+\alpha_{0}\cos^{2}\theta_{0}\\
	&+\sqrt{1-\alpha^{2}_{0}}\alpha_{p\pi^-}\sin\theta_{0}\cos\theta_{0}\sin\Delta\Phi\sin\theta_{1}\sin\phi_{1}\cos\theta_{2}\\
	&+\sqrt{1-\alpha^{2}_{0}}\alpha_{p\pi^-}\sin\theta_{0}\cos\theta_{0}\sin\Delta\Phi\cos\theta_{1}\sin\phi_{1}\sin\theta_{2}\\
        &\times\sqrt{1-\alpha^{2}_{\Lambda\pi^+}}\cos(\Delta_{\Lambda\pi^+}+\phi_{2})\\
	&+\sqrt{1-\alpha^{2}_{0}}\alpha_{p\pi^-}\sin\theta_{0}\cos\theta_{0}\sin\Delta\Phi\cos\phi_{1}\sin\theta_{2}\\
        &\times\sqrt{1-\alpha^{2}_{\Lambda\pi^+}}\sin(\Delta_{\Lambda\pi^+}+\phi_{2})\\
	&+\sqrt{1-\alpha^{2}_{0}}\alpha_{\Lambda\pi^+}\sin\theta_{0}\cos\theta_{0}\sin\Delta\Phi\sin\theta_{1}\sin\phi_{1}\\
	&+\alpha_{0}\alpha_{p\pi^-}\alpha_{\Lambda\pi^+}\cos^{2}\theta_{0}\cos\theta_{2}\\
	&+\alpha_{p\pi^-}\alpha_{\Lambda\pi^+}\cos\theta_{2},
	\end{split}
\end{equation}
where $\alpha_{p\pi^-}$ denotes the decay asymmetry parameter in the weak hadronic decay $\Lambda\to p\pi^{-}$. 
Formulas of other two-body decays can be found in Ref.~\cite{BESIII:2019odb}. 

Although the analysis of the angular distribution of these four two-body CF channels provides effective constraints, the three-body channel $\Lambda_c^+\to pK^-\pi^+$, with a low background and large production rate, brings additional sensitivity. 
Benefiting from the transition amplitude analysis reported by the LHCb Collaboration~\cite{LHCb:2023crj}, the angular distribution of the decay chain $e^+e^-\to \gamma^*\to \Bar{\Lambda}_c^-\Lambda_c^+(\to pK^-\pi^+)$ can be expressed explicitly as follows

\begin{equation}
    \label{eq:lhcb-decay-rate}
    \small
    \begin{split}
    &\frac{\text{d}\mathcal{\varGamma}}{\text{d}\cos\theta_0~\text{d}\cos{\theta_1}~\text{d}\phi_1~\text{d}\phi_2}\\
    &\propto I_{0}(\kappa)\\
    &+I_{0}(\kappa)\cos^2\theta_0\\
    &+I_{0}(\kappa)\sqrt{1-\alpha^{2}_{0}}\sin\theta_{0}\cos\theta_{0}\sin\Delta\Phi\sum_{j=1}^3 R_{2j}(\phi_1,\theta_1,\phi_2)\alpha_{j}(\kappa),
    \end{split}
\end{equation}
where $\phi_1,\theta_1,\phi_2$ are the helicity angles for the intermediate state ($K^-\pi^+$), $R_{ij}(\phi_1,\theta_1,\phi_2)$ represents the three-dimensional $Z$-$Y$-$Z$ rotation matrix and $\kappa$ denotes the Dalitz-plot variables $(M(pK^-)^2,M(K^-\pi^+)^2)$. 
The $\sum_{j=1}^{3}R_{2j}\alpha_j(\kappa)$ gives the second component of the polarimeter vector after rotation to the production frame, which couples to the $\mathcal{P}_{\Lambda_c^+}^{y}$ polarization term.
The $\Vec{\alpha}(\kappa)$ and $I_{0}(\kappa)$ are model-agnostic representations for polarization dependence of the decay rate and the total differential decay rate over $\kappa$, respectively, which are taken from the $\Lambda_c^+$ polarimetry with the amplitude model measured by the LHCb Collaboration~\cite{LHCb:2023crj,LHCb:2022sck}.

\subsection{Fit of the joint angular distributions}
In this analysis, the free parameters $\alpha_{0}$, $\sin\Delta\Phi$, $\alpha_{BP}$, and $\Delta_{BP}$, which describe the angular distributions for each c.m. energy are determined through a simultaneous unbinned maximum log-likelihood fit. 
The $\alpha_{0}$ and $\sin\Delta\Phi$ are specific to the individual energy point.
The effective likelihood function is constructed for a fixed sample size, and a nonextended maximum likelihood fit is performed. 
The effect of neglecting the Poisson fluctuation of the total yield on the estimated uncertainties is negligible for the parameters of interest.
The total effective log-likelihood function is obtained by summing the contributions from all energy points and can be expressed as
\begin{equation}
    \label{likelihood}
    (\text{ln}~\mathcal{L})^{\text{eff}}_{\text{total}}=\sum^{\text{energy}}\text{ln}~\mathcal{L}^{\text{eff}}_{\text{sig}}.
\end{equation}
The effective likelihood $\mathcal{L}^{\text{eff}}_{\text{sig}}$ corresponds to the signal-only case, which is equivalent to the likelihood constructed from data after background subtraction.

The background contribution to the log-likelihood is subtracted based on the calculated log-likelihoods using data in the sideband region $M_{\text{BC}}\in[2.250,2.270]~\text{GeV}/c^{2}$. 
For the mis-reconstructed signal events, the contribution to the log-likelihood is based on the signal MC simulation.
Both components are scaled to match the corresponding yields observed in data.
The scale factors $k_{\text{bkg}}$ and $k_{\text{mis}}$ are determined from the $M_{\text{BC}}$ fit, using the background yield $N_{\text{bkg}}$ in the sideband region and the mis-reconstructed signal yield $N_{\text{mis}}$, together with the corresponding yields obtained from the signal MC simulation.
Then $\text{ln}~\mathcal{L}^{\text{eff}}_{\text{sig}}$ can be expressed as
\begin{equation}
    \label{bkgsubtraction}
    \text{ln}~\mathcal{L}^{\text{eff}}_{\text{sig}}=K\times(\text{ln}~\mathcal{L}_{\text{data}}-k_\text{bkg}\text{ln}~\mathcal{L}_{\text{bkg}}-k_\text{mis}\text{ln}~\mathcal{L}_{\text{mis}}).
\end{equation}
Here, the subscripts ``bkg" and ``mis" correspond to the background and mis-reconstructed components, respectively.
The factor $K$ is used to correct uncertainties in unbinned effective maximum likelihood fits as suggested in Refs.~\cite{BESIII:2022udq,Langenbruch:2019nwe}, and it is introduced to account for the presence of event weights in the effective log-likelihood.
Following the prescription of Ref.~\cite{Langenbruch:2019nwe}, it can be written as
\begin{equation}
K = \frac{N_\text{data}-k_\text{bkg}N_\text{bkg}-k_\text{mis}N_\text{mis}}{N_\text{data}+k^2_\text{bkg}N_\text{bkg}+k^2_\text{mis}N_\text{mis}}.
\end{equation}
This factor effectively accounts for the underestimation of statistical uncertainty arising from background subtraction.
The likelihood function of each component can be constructed from the joint probability density function (PDF) with
\begin{equation}
    \label{likelihoodpdf}
    \mathcal{L}_{\text{data/bkg/mis}}=\prod_{i=1}^{N_{\text{data/bkg/mis}}}f_{s}(\vec\xi_{i};\vec\eta).
\end{equation}
Here, $f_{s}(\vec\xi_{i};\vec\eta)$ is the PDF of the signal process, $N_{\text{data}}$ is the number of data events and $i$ is the event index. 
The signal PDF $f_{s}(\vec\xi_{i};\vec\eta)$ is formulated as
\begin{equation}
	\label{signalPDF}
	f_{s}(\vec\xi_{i};\vec\eta)=\frac{\epsilon(\vec\xi_{i})|M(\vec\xi_{i};\vec\eta)|^{2}}{\int\epsilon(\vec\xi_{i})|M(\vec\xi_{i};\vec\eta)|^{2}\text{d}\vec\xi_{i}}~,
\end{equation}
where $\vec\xi_{i}$ denotes the kinematic observables, including the helicity angles and Dalitz-plot variables, while $\vec\eta$ denotes the free parameters to be determined. 
Note that $M(\vec\xi_{i};\vec\eta)$ is the total decay amplitude, and its squared modulus $|M(\vec\xi_{i};\vec\eta)|^2$ in the signal PDF corresponds to the decay rates in  Eqs.~\eqref{hel:formulalmdpi} and \eqref{eq:lhcb-decay-rate}. 
The function $\epsilon(\vec\xi_{i})$ is the detection efficiency, parametrized in terms of the kinematic variables $\vec\xi_{i}$. 
It is important to note that $\vec\xi_{i}$ is calculated using four-momenta updated by the kinematic fit, which constrains the invariant mass of particle decay products in the signal events to their world average masses~\cite{ParticleDataGroup:2024cfk}.
The normalization factor is calculated with a large phase-space signal MC sample as $\int\epsilon(\vec\xi_{i})|M(\vec\xi_{i};\vec\eta)|^{2}\text{d}\vec\xi_{i}=\frac{1}{N_{\text{gen}}}\sum_{k_{\text{surv}}}^{N_{\text{surv}}}|M(\vec\xi_{k_{\text{surv}}};\vec\eta)|^{2}$, where $N_{\text{gen}}$ is the total number of generated events, $N_{\text{surv}}$ is the number of events surviving all selection criteria and $k_{\text{surv}}$ is the event index.
The size of the phase-space signal simulation sample is sufficiently large compared to the statistics in data, corresponding to approximately 300 simulated events per data event. Therefore, the statistical uncertainty associated to efficiency is negligible.

The negative total effective log-likelihood with background subtraction over the five signal channels and data samples collected from 13 energy points is minimized using the \textsc{Minuit}~\cite{James:1975dr} algorithms.
The decay asymmetry parameters of the weak decays $\Lambda\to p\pi^-$ and $\Sigma^+\to p\pi^0$ and their charge conjugated processes $\bar\Lambda\to \bar p\pi^+$ and $\bar\Sigma^-\to \bar p\pi^0$ are fixed according to the experimentally measured decay asymmetry parameters~\cite{BESIII:2022qax,ParticleDataGroup:2024cfk}.
In the nominal fit, the parameters $\alpha_{BP}$ and $\Delta_{BP}$ are set to $-\bar\alpha_{BP}$ and $-\bar\Delta_{BP}$, respectively, in accordance with the assumption of $C\!P$ conservation. 
Here, $\bar\alpha_{BP}$ and $\bar\Delta_{BP}$ represent the parameters associated with the antiparticle $\bar\Lambda_c^-$ decays.
The results of the fit are presented in Fig.~\ref{fig:fitangles}, and the full correlation matrix is provided in Table~\ref{Res:CPCcov} of Appendix~\ref{App:A}. 
An alternative test is also performed where the assumption of $C\!P$ conservation is relaxed and $\bar\alpha_{BP}$ and $\bar\Delta_{BP}$ are treated as free parameters in the fit. 

\begin{figure*}[htbp]
	\centering
	\includegraphics[width=\linewidth]{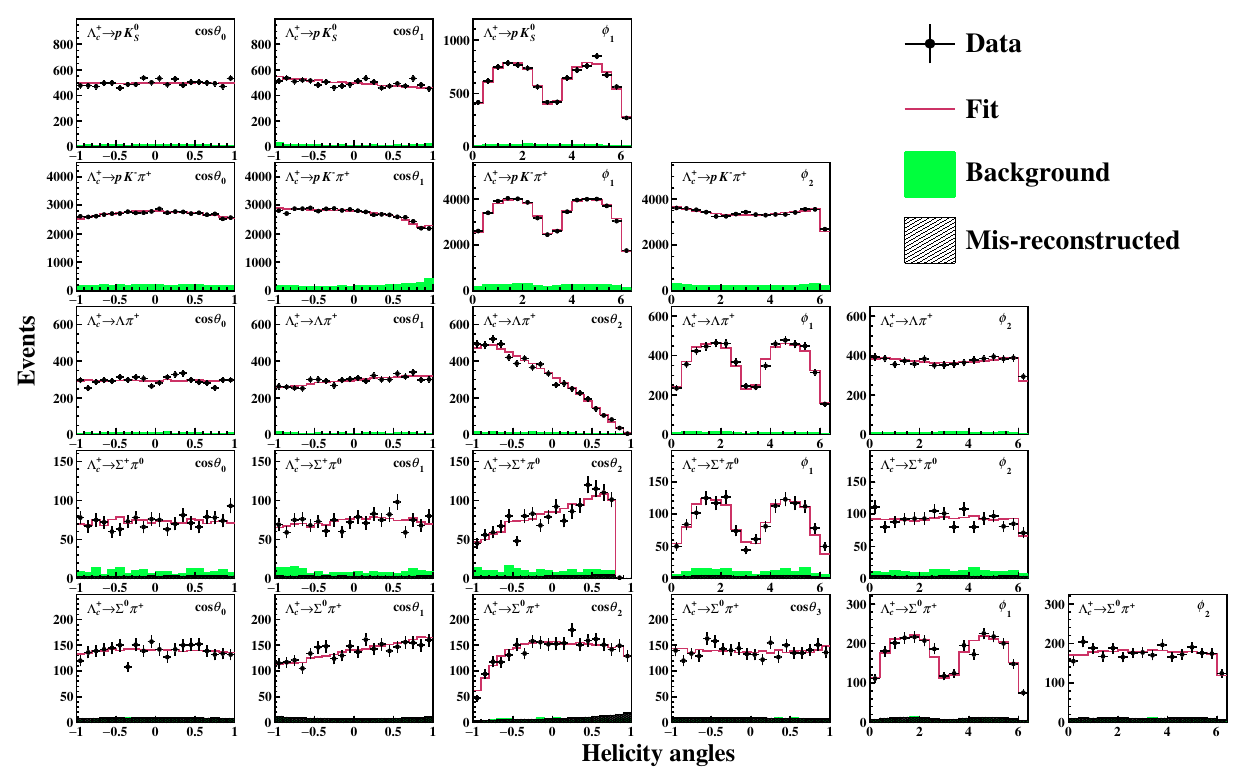}
	\caption{Fit results of helicity angles combining all energy points. Shown are angular distributions about $pK_S^0$, $pK^-\pi^+$, $\Lambda\pi^+$, $\Sigma^+\pi^0$, and $\Sigma^0\pi^+$, respectively. Black points with error bars are data. The green shaded region indicates combinatorial background events and black shaded region shows the mis-reconstructed signal events.}
	\label{fig:fitangles}
\end{figure*}

\subsection{Systematic uncertainties}

The systematic uncertainties originate from the differences between simulations and real physical processes leading to deviations in detection efficiencies. 
Here, the absolute systematic uncertainties are estimated. 
The numerical values are summarized in Table~\ref{ParaAna:SysUnALL}, and the individual sources are discussed in the rest of this section.

\begin{itemize}
\item \textit{Selection criteria.} The systematic uncertainties regarding selection criteria primarily arise from the reconstruction of final states, $\Delta E$ requirement, and $M_{\text{BC}}$ signal selection. 
In general, the tracking and PID efficiency difference between data and simulation samples can be investigated from the control samples. 
Hence, the efficiency difference in each $\cos\theta$ and transverse momentum bin is applied to correct the phase-space MC sample. 
The decay parameters are then reevaluated using the corrected MC samples, and the resulting differences from the nominal values are taken as the corresponding systematic uncertainties.
Specifically, the reconstruction efficiency of proton, pion, and kaon tracks is studied with control samples of $e^{+}e^{-}\to p\bar{p}\pi^{+}\pi^{-}$, $e^{+}e^{-}\to\pi^{+}\pi^{-}\pi^{+}\pi^{-}$, and $J/\psi\to K^{0}_{S}K^{\pm}\pi^{\mp}$ decays, respectively. 
The $\pi^{0}$ meson reconstruction efficiency is studied with the $\psi(3686)\to\pi^{0}\pi^{0}J/\psi$ and $e^{+}e^{-}\to\omega\pi^{0}$ decays. 
The control channels for the $K^{0}_{S}$ reconstruction efficiency are the $J/\psi\to\phi K^{0}_{S}K^\mp\pi^\pm$ and $J/\psi\to K^*(892)^\pm K^\mp$ decays. 
For the reconstruction efficiency of the $\Lambda$ baryon the $J/\psi\to\bar{p}K^{+}\Lambda$ and $J/\psi\to\Lambda\bar{\Lambda}$ decay samples are used~\cite{BESIII:2018ciw}. 
The reconstruction efficiency for photons is studied using the $e^+e^-\to\gamma\mu^+\mu^-$ decay and small discrepancies between data and MC simulation are observed, amounting to $0.5\%$ and $1.5\%$ in the barrel and end cap of the EMC. 
A randomly generated efficiency correction factor is used to adjust the phase-space simulation. 
This procedure introduces the fluctuation reflecting the observed data–MC efficiency difference.
Requirements relating to the MUC penetration depth, probability of the electron, and pion assumption are studied with a control sample of $\Lambda^{+}_{c}\to\Lambda\pi^{+}$ decays. 
The potential differences in event selection efficiencies for data and MC simulation are taken into account in the fit described in the previous section via reweighting the phase-space sample. 
The uncertainties related to the $\Delta E$ and $M_{\text{BC}}$ requirements are evaluated by smearing the signal simulation with a Gaussian resolution function. 
The changes in the fit results based on smeared signal simulation are taken to be systematic uncertainties.
\item \textit{Performance of matching algorithm.} In the nominal measurement, the angle between the reconstructed and true photon momentum directions from simulations is required to be below 10 degrees. 
To evaluate the effect due to this matching algorithm, two different matching conditions of 9 and 11 degrees are tested and the resultant largest difference from the nominal result is considered as the systematic uncertainty.
\item \textit{Signal components.} The amplitude-model related uncertainties are computed using the alternative models in Ref.~\cite{LHCb:2022sck}. 
With alternative $\Lambda^{+}_{c}$ polarimetry, the angular fit is repeated and the standard deviations and difference of mean value are assigned to the model uncertainties. 
In Ref.~\cite{LHCb:2023crj}, the statistical uncertainties are estimated by varying the parameter values in the default model. 
These parameters are sampled from a Gaussian function, whose $\mu$ and $\sigma$ are given by the central and error values from the default model fit results in Ref.~\cite{LHCb:2022sck}. 
The polarimeter vector fields are reproduced for each varied parameter set. 
The angular fit procedure is repeated using the corresponding inputs, and the statistical component is given by taking the standard deviations and the difference of the mean value.
\item \textit{Background components.} Background fractions are fixed according to the fit results from the nominal fit. 
To estimate systematic uncertainties, the background fraction for each energy point is varied by $\pm 1\sigma$ and the fit procedure is repeated. 
The quadratic sum of the largest variation from each fraction is assigned as a systematic uncertainty from the background fraction. 
The mis-reconstructed model is examined using an alternative signal simulation produced with new input parameters $\alpha_{\Sigma^{0}\pi^{+}}$ and $\alpha_{\Sigma^{+}\pi^{0}}$ which are changed within $\pm 1\sigma$ of the results presented here.
The background shape is extracted from the data events in the $M_{\text{BC}}$ sideband regions. 
For the $\Lambda_c^+\to pK^-\pi^+$ channel, to estimate the possible difference between the $M_{\text{BC}}$ sideband and signal regions, a weight factor is derived by comparing the $M(K^-\pi^+)^2$ distributions in the $M_{\text{BC}}$ sideband and signal regions of inclusive simulation, since the most significant discrepancy appears in the $M(K^-\pi^+)^2$ distribution, while the others show consistent behavior. 
Data events in the $M_{\text{BC}}$ sideband region are corrected by the weight factor and the fit procedure is repeated. 
The difference is then taken as the uncertainty from the background model. 
For two-body decays, the uncertainty due to the combinational background model is estimated by varying the relative weights between $\Lambda_{c}^{+}\bar{\Lambda}_{c}^{-}$ pairs and other hadronic events based on the uncertainties of their cross-section ratio. 
The relevant systematic uncertainties from background size are examined by repeating the fits with an alternative background size obtained from the Gaussian sampling of the fitted parameters.
The series of fitted parameters, assumed to follow a Gaussian distribution, are used to estimate the systematic uncertainty for the background size. 
The fitted Gaussian means are found to be consistent with the nominal result, typically with central value deviating at the level of $10^{-3}$. 
To be conservative, the corresponding systematic uncertainty is determined by summing the fitted Gaussian width and the deviation of the fitted Gaussian mean from the nominal value.

\item \textit{Fixed parameters.} The systematic uncertainty due to the input parameters is evaluated by varying these parameters using a Gaussian sampling method. 
For each parameter, the obtained results are expected to follow Gaussian distributions. 
The sum of the fitted Gaussian resolution and the difference between the fitted Gaussian mean and the nominal result is taken as the systematic uncertainty.
The fitted means are close to zero~\cite{Zhong:2024qqs}, but they are included with the width as a conservative estimation.
\end{itemize}

\begin{table}[htbp]
	\centering
	\small
	\caption{Absolute systematic uncertainties in transverse polarization and decay asymmetry parameters~(in percent), where SEL, PMA, SIG, BKG, and PAR in the table represent the sources from selection criteria, Performance of the matching algorithm, signal components, background components, and fixed parameters, respectively. Some negligible terms are denoted by ``$\raisebox{0.5ex}{...}$''. The $\alpha_{0}$ and $\sin\Delta\Phi$ are given as ranges because they vary for different energy points. The total uncertainty is calculated as the sum in quadrature of all components.}
	\label{ParaAna:SysUnALL}
	\renewcommand{\arraystretch}{1.5}
	\begin{tabular}{l c c c c c c}
	\hline\hline
        Parameters\qquad & SEL & PMA & SIG & BKG & PAR & Total \\
	\hline
$\alpha_{0}$           & $\le0.5$ & \raisebox{0.5ex}{...} & $\le0.3$ & $0.1-2.0$ & \raisebox{0.5ex}{...} & $0.1-2.0$ \\
$\sin\Delta\Phi$           & $\le0.6$ & \raisebox{0.5ex}{...} & $0.5-4.8$ & $0.2-2.9$ & $\le0.3$ & $0.9-5.0$ \\
$\langle\alpha_{pK^{0}_{S}}\rangle$          & $0.1$ & \raisebox{0.5ex}{...} & $1.1$ & $2.8$ & $0.1$ & $3.0$ \\
$\langle\alpha_{\Lambda\pi^{+}}\rangle$      & $0.8$ & \raisebox{0.5ex}{...} & \raisebox{0.5ex}{...} & $0.6$ & $0.3$ & $1.1$ \\
$\langle\Delta_{\Lambda\pi^{+}}\rangle$      & $0.7$ & \raisebox{0.5ex}{...} & $0.8$ & $1.1$ & $0.1$ & $1.5$ \\
$\langle\alpha_{\Sigma^{0}\pi^{+}}\rangle$   & $0.3$ & \raisebox{0.5ex}{...} & $0.1$ & $0.9$ & $0.2$ & $1.0$ \\
$\langle\Delta_{\Sigma^{0}\pi^{+}}\rangle$   & $0.8$ & \raisebox{0.5ex}{...} & $1.1$ & $2.3$ & $0.3$ & $2.7$ \\
$\langle\alpha_{\Sigma^{+}\pi^{0}}\rangle$   & $1.2$ & \raisebox{0.5ex}{...} & \raisebox{0.5ex}{...} & $1.5$ & $1.3$ & $2.3$ \\
$\langle\Delta_{\Sigma^{+}\pi^{0}}\rangle$   & $1.5$ & \raisebox{0.5ex}{...} & $1.2$ & $3.9$ & $0.4$ & $4.4$ \\
    \hline\hline    
	\end{tabular}
\end{table}

\section{Results}

\subsection{Transverse polarization and form factors}
\begin{figure}[htbp]
	\centering
	\includegraphics[width=\linewidth]{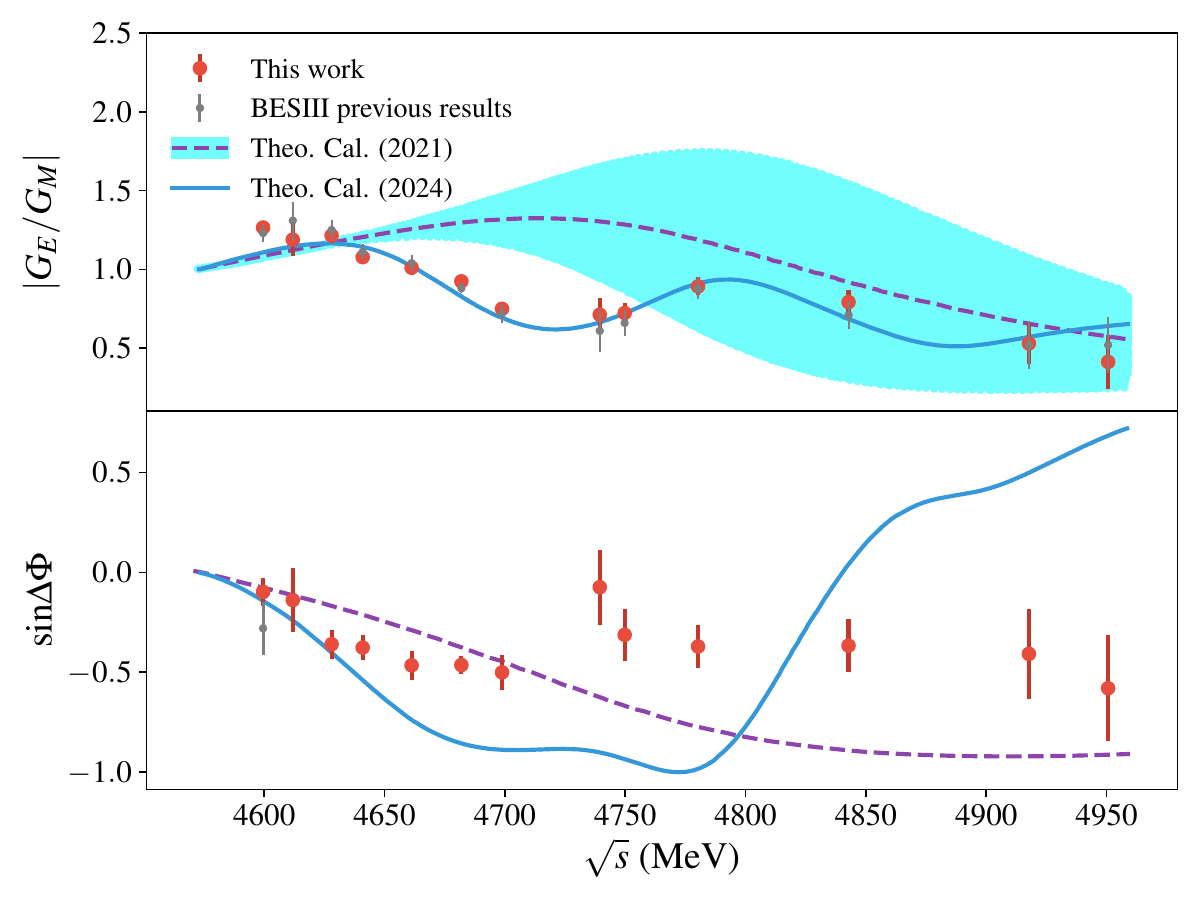}
	\caption{Comparison of $|G_E/G_M|$ and $\sin\Delta\Phi$ between theoretical calculations and this work. Red points with error bars correspond to this work, and the gray ones are previous measurement results from BESIII~\cite{BESIII:2019odb,BESIII:2023rwv}. The purple dashed line with sky-blue shading is calculated from  Ref.~\cite{Wan:2021ncg}, and the blue solid line represents the calculated results with Ref.~\cite{Chen:2023oqs}.}
	\label{fig:sindelta0}
\end{figure}

The results of the angular distribution fit, the $\alpha_0$, the sine of the phase difference $\sin\Delta\Phi$, its significance, and the rms value of transverse polarization $\boldsymbol{\sqrt{\langle{|\mathcal{P}_{\Lambda_c^+}^{y}|}^2\rangle}}$, are summarized in Table~\ref{tab:PT}.  
As discussed in Ref.~\cite{Salone:2022lpt}, the average transverse polarization squared 
$\boldsymbol{\langle{|\mathcal{P}_{\Lambda_c^+}^{y}|}^2\rangle}$ can be calculated using the integration $\frac{3}{2}\int{\boldsymbol{{|\mathcal{P}_{\Lambda_c^+}^{y}|}^2}\frac{1+\alpha_0\cos^2\theta_0}{3+\alpha_0}}\text{d}\cos\theta_0$,
where $\boldsymbol{\mathcal{P}_{\Lambda_c^+}^{y}}$ is given by Eq.~\eqref{hel:pt}.
The measured $\alpha_0$ is related to the electric and magnetic form factors $G_E$ and $G_M$ via $\alpha_0=(|G_E|^2(1-v^2)-|G_M|^2)/(|G_E|^2(1-v^2)+|G_M|^2)$,
where $v$ is the $\Lambda_c^+$ velocity in the c.m. system, normalized to the speed of light. 
Our result for $\alpha_0$ is consistent with previous measurements and enables an updated determination of the ratios $|G_E/G_M|$, which is compared to previous BESIII measurements~\cite{Chen:2023oqs} and theoretical calculations~\cite{Wan:2021ncg,Chen:2023oqs}, as shown in Fig.~\ref{fig:sindelta0}. 
The measurements of the $q^2$ dependence of the sine of the form factor phase, $\sin\Delta\Phi$, are also presented. The 13 data points provide the most fine-grained determination of the $q^2$-dependence of $\sin\Delta\Phi$ for any baryon so far: Previous measurements for $\Lambda$ and $\Sigma^+$ hyperons contain five \cite{BESIII:2025kyg} and three \cite{BESIII:2023ynq} data points. Theoretical calculations based on vector meson dominance (VMD) ~\cite{Wan:2021ncg,Chen:2023oqs} have been performed. Our data differ significantly from these predictions, revealing a more complicated structure.

\begin{table*}[htbp]
	\small
	\centering
	\caption{Results for $\alpha_0$, $\sin\Delta\Phi$, and transverse polarization of $\Lambda_{c}^{+}$ at 13 energy points, where the first uncertainties are statistical and the second are systematic. In the table, $SL$ means the significance level of nonzero $\sin\Delta\Phi$, $\sigma$ represents the standard deviation, and $\boldsymbol{\sqrt{\langle{|\mathcal{P}_{\Lambda_c^+}^{y}|}^2\rangle}}$ is the rms value of transverse polarization.}
	\label{tab:PT}
	\begin{tabular}{c c c c c c c}
	\hline\hline
        \raisebox{0.2ex}{$\sqrt{s}~[\text{MeV}]$} & \raisebox{0.2ex}{$\alpha_{0}~$in previous work~\cite{BESIII:2019odb,BESIII:2023rwv}} & \raisebox{0.2ex}{$\alpha_{0}~$in this work} & \raisebox{0.2ex}{$\sin\Delta\Phi$} & \raisebox{0.2ex}{$SL$} & \raisebox{0.5ex}{$\boldsymbol{\sqrt{\langle{|\mathcal{P}_{\Lambda_c^+}^{y}|}^2\rangle}}(\%)$} \\ 
        \hline
        $4600$ & $-0.20\pm0.04\pm0.02$ & $-0.226\pm0.030\pm0.004$ & $-0.098\pm0.069\pm0.009$ & $ 2.2\sigma$ & $3.81^{+2.66}_{-2.41}\pm0.23$ \\
        $4612$ & $-0.26\pm0.09\pm0.01$ & $-0.163\pm0.082\pm0.012$ & $-0.139\pm0.160\pm0.033$ & $ 1.1\sigma$ & $5.34^{+5.84}_{-3.68}\pm0.92$ \\
        $4628$ & $-0.21\pm0.04\pm0.01$ & $-0.181\pm0.038\pm0.001$ & $-0.368\pm0.077\pm0.011$ & $ 6.8\sigma$ & $14.19\pm2.96\pm0.42$ \\
        $4641$ & $-0.09\pm0.05\pm0.01$ & $-0.060\pm0.039\pm0.004$ & $-0.386\pm0.067\pm0.014$ & $ 7.5\sigma$ & $14.41\pm2.49\pm0.34$ \\
        $4661$ & $-0.02\pm0.05\pm0.01$ & $ 0.008\pm0.044\pm0.003$ & $-0.484\pm0.079\pm0.018$ & $ 8.6\sigma$ & $17.63\pm2.85\pm0.22$ \\
        $4682$ & $ 0.15\pm0.03\pm0.01$ & $ 0.102\pm0.029\pm0.003$ & $-0.482\pm0.048\pm0.019$ & $14.1\sigma$ & $16.86\pm1.63\pm0.28$ \\
        $4699$ & $ 0.34\pm0.07\pm0.01$ & $ 0.305\pm0.055\pm0.010$ & $-0.524\pm0.098\pm0.025$ & $ 7.2\sigma$ & $16.35\pm2.98\pm0.37$ \\
        $4740$ & $ 0.49\pm0.16\pm0.03$ & $ 0.358\pm0.126\pm0.009$ & $-0.074\pm0.189\pm0.015$ & $ 0.3\sigma$ & $2.22^{+5.10}_{-1.52}\pm0.45$ \\
        $4750$ & $ 0.42\pm0.10\pm0.01$ & $ 0.347\pm0.079\pm0.005$ & $-0.318\pm0.135\pm0.018$ & $ 3.2\sigma$ & $9.64^{+4.02}_{-4.04}\pm0.51$ \\
        $4781$ & $ 0.17\pm0.07\pm0.01$ & $ 0.157\pm0.062\pm0.008$ & $-0.380\pm0.114\pm0.025$ & $ 5.1\sigma$ & $12.94^{+3.79}_{-3.86}\pm0.44$ \\
        $4843$ & $ 0.38\pm0.10\pm0.01$ & $ 0.282\pm0.089\pm0.020$ & $-0.375\pm0.140\pm0.032$ & $ 3.6\sigma$ & $11.87^{+4.35}_{-4.39}\pm0.91$ \\
        $4918$ & $ 0.62\pm0.17\pm0.01$ & $ 0.609\pm0.152\pm0.019$ & $-0.420\pm0.248\pm0.022$ & $ 1.9\sigma$ & $9.95^{+5.33}_{-5.55}\pm0.34$ \\
        $4951$ & $ 0.63\pm0.21\pm0.01$ & $ 0.746\pm0.183\pm0.009$ & $-0.618\pm0.326\pm0.050$ & $ 1.7\sigma$ & $11.81^{+6.10}_{-6.72}\pm0.97$ \\
        \hline\hline
	\end{tabular}
\end{table*}

\begin{figure}[htbp]
	\centering
	\includegraphics[width=\linewidth]{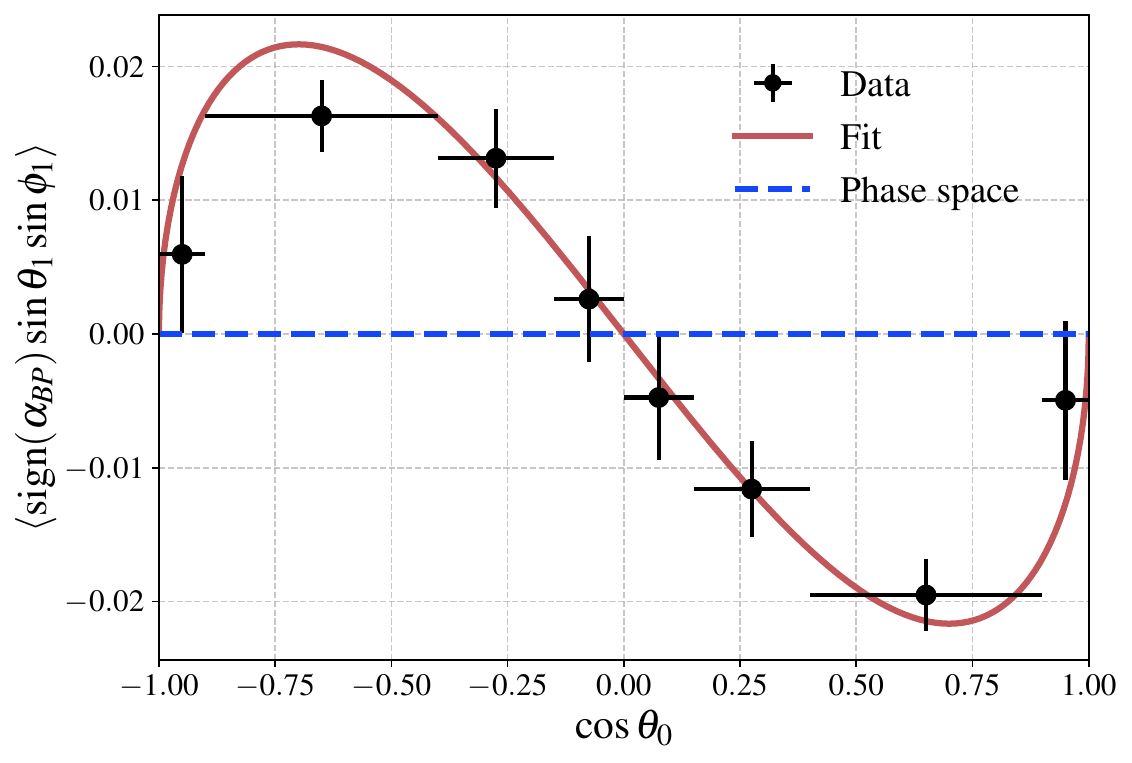}
	\caption{ Moment $\textless\text{sign}(\alpha_{BP})\sin\theta_{1}\sin\phi_{1}\textgreater$ constructed to illustrate the effect of transverse polarization, which is obtained by averaging over all data samples at different energy points. Points with error bars correspond to data with background subtracted and the red solid line denotes the global expected dependence on $\cos\theta_{0}$. The blue dashed line represents the distribution without polarization uniformly distributed in phase space. }
	\label{fig:momentcolor}
\end{figure}

Of particular interest is the fact that the $\Lambda_c^+$ baryon is significantly polarized with respect to the normal of the production plane. 
The rms value of transverse polarization reaches a maximum of approximately 18\% at 4661 MeV.
To illustrate the polarization effect, the moment $\textless\sin\theta_{1}\sin\phi_{1}\textgreater$ is calculated as the average of $\sin\theta_{1}\sin\phi_{1}$ within a certain range of $\cos\theta_0$. 
It is expected to be proportional to $\alpha_{BP}\boldsymbol{\mathcal{P}_{\Lambda_c^+}^{y}}(\cos\theta_{0})/(1+\alpha_{0} \cos^{2}\theta_0)$. 
We divide the range of $\cos\theta_0$, $(-1,1)$, into eight bins, calculate the moment for experimental data, and provide the best-fit curve at all energy points, as shown in Fig.~\ref{fig:momentcolor}. 
Fits are performed with $\sin\Delta\Phi$ either left free to float or fixed at zero to obtain the corresponding likelihood values. 
The difference in likelihood between these two scenarios and the change in the number of degrees of freedom are then used to evaluate the statistical significance of a nonzero $\sin\Delta\Phi$.
The sign($\alpha_{BP}$) is used to avoid cancellation between charge-conjugated processes. 
The significance is reported in Table~\ref{tab:PT}.

\subsection{Decay asymmetry and the SM test}
\begin{figure*}[htbp]
	\centering
        \includegraphics[width=\linewidth]{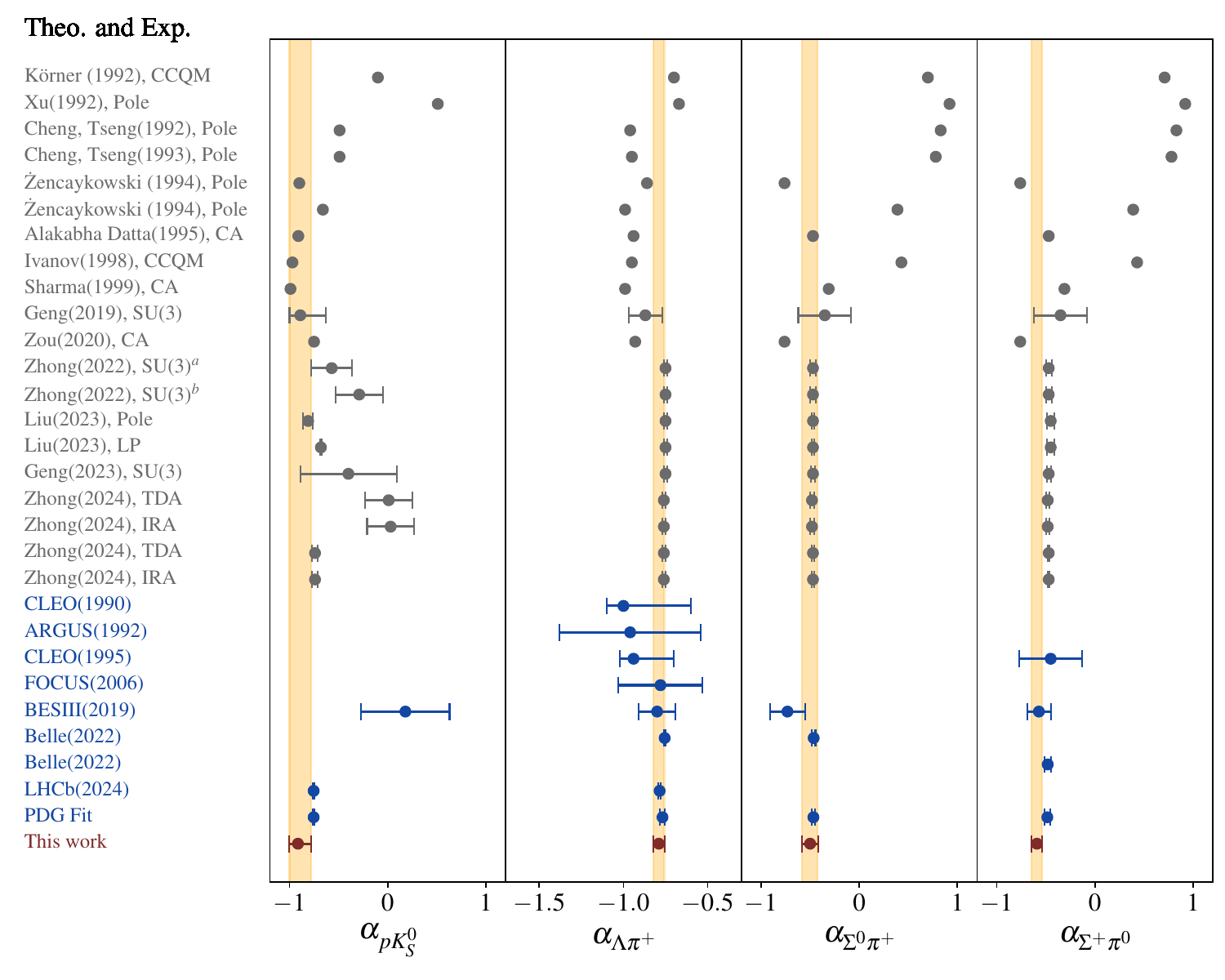}
	\caption{Comparison about theoretical calculations and experimental measurements of decay asymmetry parameters. Gray points are theoretical calculations~\cite{Korner:1992wi,Xu:1992vc,Cheng:1991sn,Cheng:1993gf,Zenczykowski:1993hw,Zenczykowski:1993jm,Datta:1995mn,Ivanov:1997ra,Sharma:1998rd,Geng:2019xbo,Zou:2019kzq,Zhong:2022exp,Liu:2023dvg,Geng:2023pkr,Zhong:2024zme,Zhong:2024qqs,Cheng:2024lsn}, and experimental measurements~\cite{CLEO:1990unw,ARGUS:1991yzs,CLEO:1995qyd,FOCUS:2005vxq,BESIII:2019odb,Belle:2022uod,Belle:2022bsi,LHCb:2024tnq,ParticleDataGroup:2024cfk} are marked in blue. Red points and yellow shadows represent the results obtained in this work. In the legend, the names of the theoretical studies and experimental measurements match the colors used in the figures, arranged in chronological order. For each theoretical study, the main theoretical models used are also specified.}
	\label{fig:band}
\end{figure*}

As the SM predicts that $C\!P$ violation is highly suppressed in these decay channels, we assume $C\!P$ symmetry, meaning the decay asymmetry parameters $\alpha_{BP}=-\bar{\alpha}_{BP}$ and $\Delta_{BP}=-\bar{\Delta}_{BP}$ in the angular distribution analysis. 
In addition to the transverse polarization, the decay asymmetry parameters $\langle\alpha_{BP}\rangle$ and $\langle\Delta_{BP}\rangle$ are directly obtained in the fitting and shown in Table~\ref{tab:Asycpc} and Fig.~\ref{fig:band} for numerical results and comparison with theoretical calculations, respectively. 
Only the parameter $\langle\alpha_{BP}\rangle$ is extracted for $\Lambda_{c}^{+}\to pK^{0}_{S}$ since no information on the proton spin orientation is available in the present experiment. 
Then, $\langle\beta_{BP}\rangle$ and $\langle\gamma_{BP}\rangle$ can be derived using $\langle\alpha_{BP}\rangle$ and $\langle\Delta_{BP}\rangle$ for $\Lambda_{c}^{+}\to\Lambda\pi^+$, $\Lambda_{c}^{+}\to\Sigma^0\pi^+$, and $\Lambda_{c}^{+}\to\Sigma^+\pi^0$ decays, which are given in Table~\ref{tab:Asycpc}.

\begin{figure*}[htbp]
	\centering
        \includegraphics[width=\linewidth]{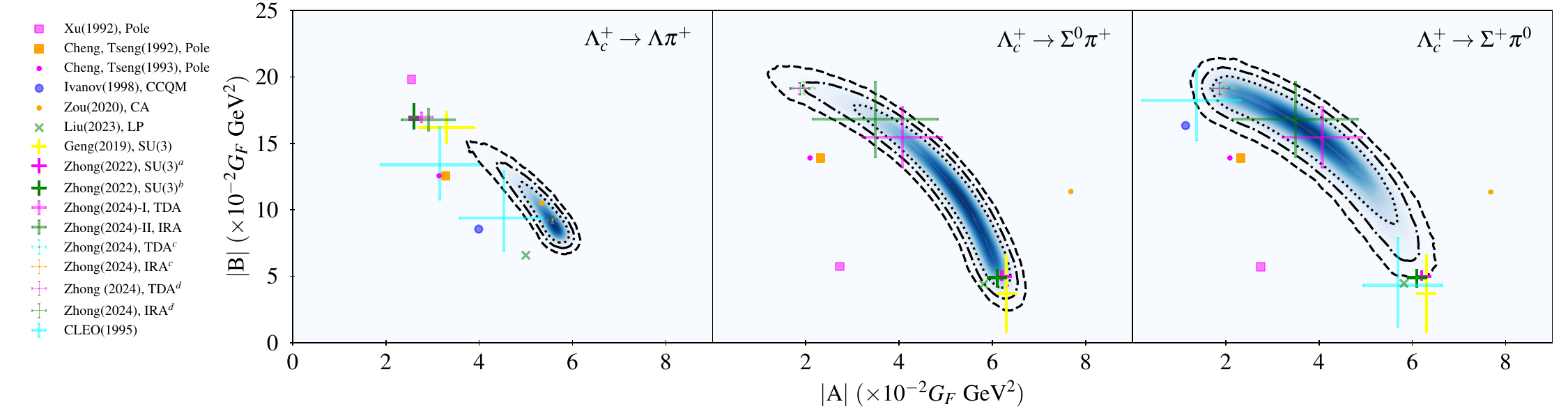}
	\caption{Two-dimensional density of $|A|$ and $|B|$. Markers are theoretical calculations~\cite{Xu:1992vc,Cheng:1991sn,Cheng:1993gf,Ivanov:1997ra,Zou:2019kzq,Liu:2023dvg,Geng:2019xbo,Zhong:2022exp,Zhong:2024qqs,Cheng:2024lsn} and experimental measurements~\cite{CLEO:1995qyd}, and shadows representing the distribution of $|A|$ and $|B|$ obtained in this work; $G_F$ is the Fermi coupling constant; and $c$ and $d$ denote whether the model uses the results from the Belle experiment with the BFs and decay asymmetry parameters of $\Xi_c^{0}\to\Xi^0\pi^0$, $\Xi^0\eta$, and $\Xi^0\eta'$. The contours correspond to 68.2\%, 95.4\%, and 99.7\% confidence levels, respectively.}
	\label{fig:AB2d}
\end{figure*}

Since the information on $|A|$ and $|B|$ in $\Lambda_{c}^{+}\to\Xi^{0}K^{+}$~\cite{BESIII:2023wrw} has proven useful for the development of effective models~\cite{Zhong:2024zme,Zhong:2024qqs,Cheng:2024lsn}, we also derive these amplitudes in this work.  The corresponding two-dimensional density distributions are shown in Fig.~\ref{fig:AB2d}. We do not quote numerical results as the resulting contours exhibit nontrivial shapes that cannot be adequately summarized by simple confidence intervals.

A secondary fit is also performed where the $C\!P$ constraint is removed and is used to test the SM, and the results are shown in Appendix~\ref{App:B}. 
The $\alpha$-induced $C\!P$ observable and weak and strong phase differences are constructed in $A^{\alpha_{BP}}_{C\!P}$, $\text{tan}\phi_{C\!P}$, and $\text{tan}\Delta_{S}$.
The first two observables, $A^{\alpha_{BP}}_{C\!P}$ and $\tan\phi_{C\!P}$, are $C\!P$-violating observables, corresponding to the two constraints $\alpha_{BP}=-\bar{\alpha}_{BP}$ and $\Delta_{BP}=-\bar{\Delta}_{BP}$ imposed on the $C\!P$-conserving baseline fit, while $\tan\Delta_{S}$ is a $C\!P$-conserving observable.
The $\phi_{C\!P}$ and $\Delta_{S}$ denote the weak and strong interaction phase differences between the $P$ wave and the $S$ wave.
The results are summarized in Table~\ref{tab:Asycpv}.
Some of the parameters in the table are derived from the fit results, and their uncertainties account for correlations among the fit parameters. 
The full correlation matrix is provided in Table~\ref{Res:CPVcov} in Appendix~\ref{App:A}.

\begin{table*}[htbp]
	\centering
	\caption{Results of the decay observables, where all results are obtained based on the assumption of $C\!P$ conservation. Terms that could not be extracted are denoted by ``$\raisebox{0.5ex}{...}$''. All angular values in the table are expressed in radians. Only the $\langle\alpha_{BP}\rangle$ and $\langle\Delta_{BP}\rangle$ are directly obtained from the fit, and $\langle\beta_{BP}\rangle$, $\langle\gamma_{BP}\rangle$, $\langle\Delta_S\rangle$, and $\tan\langle\Delta_S\rangle$ are observables derived from them.}
	\label{tab:Asycpc}
    \begin{tabular}{c l l l l}
        \hline\hline
        \raisebox{-0.2ex}{Parameter} & \raisebox{-0.2ex}{$\phantom{-}\Lambda_c^+\to pK_S^0$} & \raisebox{-0.2ex}{$\phantom{-}\Lambda_c^+\to\Lambda\pi^+$} & \raisebox{-0.2ex}{$\phantom{-}\Lambda_c^+\to\Sigma^0\pi^+$} & \raisebox{-0.2ex}{$\phantom{-}\Lambda_c^+\to\Sigma^+\pi^0$} \\ 
        \midrule
        $\langle\alpha_{BP}\rangle$      & $-0.914^{+0.132}_{-0.086}\pm0.030$        & $-0.789\pm0.032\pm0.011$                       & $-0.499\pm0.079\pm0.010$                        & $-0.590\pm0.049\pm0.023$ \\
        $\langle\Delta_{BP}\rangle$      & $\phantom{-}$\raisebox{0.5ex}{...}        & $\phantom{-}0.638\pm0.444\pm0.015$             & $\phantom{-}1.083\pm0.634\pm0.027$              & $\phantom{-}1.898\pm0.602\pm0.044$ \\
        \midrule
        $\langle\beta_{BP}\rangle$       & $\phantom{-}$\raisebox{0.5ex}{...}        & $\phantom{-}0.366^{+0.173}_{-0.246}\pm0.011$   & $\phantom{-}0.766^{+0.084}_{-0.352}\pm0.012$    & $\phantom{-}0.764^{+0.051}_{-0.236}\pm0.019$ \\
        $\langle\gamma_{BP}\rangle$      & $\phantom{-}$\raisebox{0.5ex}{...}        & $\phantom{-}0.493^{+0.103}_{-0.202}\pm0.013$   & $\phantom{-}0.406^{+0.358}_{-0.528}\pm0.021$    & $-0.260^{+0.477}_{-0.384}\pm0.030$ \\
        $\langle\Delta_S\rangle$         & $\phantom{-}$\raisebox{0.5ex}{...}        & $\phantom{-}2.71^{+0.28}_{-0.17}\pm0.02$       & $\phantom{-}2.15^{+0.30}_{-0.09}\pm0.02$        & $\phantom{-}2.23^{+0.19}_{-0.06}\pm0.03$ \\
        $\tan\langle\Delta_{S}\rangle$   & $\phantom{-}$\raisebox{0.5ex}{...}        & $-0.46^{+0.29}_{-0.21}\pm0.02$                 & $-1.53^{+0.68}_{-0.42}\pm0.05$                  & $-1.29^{+0.40}_{-0.18}\pm0.08$ \\
        \hline\hline
	\end{tabular}
\end{table*}

\begin{table*}[htbp]
	\centering
	\caption{Results of decay observables, where $A^{\alpha_{BP}}_{C\!P}$,  $\text{tan}\phi_{C\!P}$
    , and $\text{tan}\Delta_{S}$ 
    are obtained under the assumption of $C\!P$ violation. Terms that could not be extracted are denoted by ``$\raisebox{0.5ex}{...}$''. The $A^{\alpha_{BP}}_{C\!P}$, $\text{tan}\phi_{C\!P}$, and $\text{tan}\Delta_{S}$ 
    are derived from the results given in Appendix~\ref{App:B}.}
	\label{tab:Asycpv}
    \begin{tabular}{c l l l l}
        \hline\hline
        \raisebox{-0.2ex}{Parameter} & \raisebox{-0.2ex}{$\phantom{-}\Lambda_c^+\to pK_S^0$} & \raisebox{-0.2ex}{$\phantom{-}\Lambda_c^+\to\Lambda\pi^+$} & \raisebox{-0.2ex}{$\phantom{-}\Lambda_c^+\to\Sigma^0\pi^+$} & \raisebox{-0.2ex}{$\phantom{-}\Lambda_c^+\to\Sigma^+\pi^0$} \\ 
        \midrule
        $A^{\alpha_{BP}}_{C\!P}$    & $\phantom{-}0.097^{+0.113}_{-0.096}\pm0.025$ & $\phantom{-}0.006\pm0.048\pm0.016$          & $\phantom{-}0.206^{+0.185}_{-0.147}\pm0.023$    & $-0.087\pm0.087\pm0.021$ \\
        $\text{tan}\phi_{C\!P}$     & $\phantom{-}$\raisebox{0.5ex}{...}        & $\phantom{-}0.231\pm0.242\pm0.023$             & $\phantom{-}0.592\pm0.632\pm0.036$              & $0.026\pm0.492\pm0.032$ \\
        $\text{tan}\Delta_{S}$      & $\phantom{-}$\raisebox{0.5ex}{...}        & $-0.488\pm0.242\pm0.027$                       & $-1.171\pm0.646\pm0.045$                        & $-1.272\pm0.496\pm0.049$ \\
        \hline\hline
	\end{tabular}
\end{table*}

\section{Discussion}

Transverse polarization of $\Lambda_{c}^{+}$ baryons produced in $e^+e^-$ annihilation is observed for the first time from $6.4~\text{fb}^{-1}$ of data collected at 13 c.m. energies between $4600$~MeV and $4951$~MeV with the BESIII detector. 
The observed polarization is induced by a nonzero, relative phase between the electric and the magnetic form factors of the $\Lambda^+_c$, reflecting complex dynamics in its electromagnetic structure. These data provide the currently most fine-grained determination of the $q^2$ dependence of the phase for any baryon. The results exhibit clear deviations from theoretical predictions based on vector meson dominance, revealing that a more complex inner structure is at play.

The VMD model offers a framework to describe how the production cross sections of baryon-antibaryon pairs via electron-positron annihilations change with energy~\cite{Li:2021lvs,Chen:2023oqs,Chen:2024luh}, assuming that the initial electromagnetic process dominates, with the virtual photon coupling to the produced hadrons primarily through intermediate vector mesons.
This approach has also been successfully applied to describe nucleon form factors~\cite{Yang:2019mzq}.
In 2021, a VMD model incorporating the vector charmonium states was proposed~\cite{Wan:2021ncg} to describe the Born cross sections of $e^+e^-\to\Lambda_c^+\bar{\Lambda}_c^-$ measured from Belle~\cite{Belle:2008xmh} and BESIII~\cite{BESIII:2017kqg}.
At that time, Belle data determined the overall line shape, and the form factors $|G_E/G_M|$ in the VMD model were less constrained with sparse BESIII data.
Two years later, BESIII reported the cross sections and  the form factor ratio $|G_E/G_M|$ with much improved precision at more energy points~\cite{BESIII:2023rwv}, significantly constraining the line shape.
A subsequent VMD-based study including higher vector states such as $\psi(4500)$, $\psi(4660)$, $\psi(4790)$, and $\psi(4900)$ is derived to achieve an excellent description of the cross sections and $|G_E/G_M|$ at different energy points~\cite{Chen:2023oqs}.
Unfortunately, neither calculation~\cite{Wan:2021ncg,Chen:2023oqs} reproduces the measured phase difference between $G_E$ and $G_M$, indicating a persistent discrepancy between theory and experiment.
This finding suggests that the internal structure of the $\Lambda_c^+$ and its production dynamics in $e^+e^-$ annihilation are not yet fully understood.

The observed transverse polarization enables precise measurements of the $\Lambda_c^+$ decay asymmetry parameters~\cite{Li:2025nzx}.
The key two-body decays $\Lambda_{c}^{+}\to pK^{0}_{S}$, $\Lambda\pi^{+}$, $\Sigma^{0}\pi^{+}$, and $\Sigma^{+}\pi^{0}$ have been studied.
The measured $\alpha_{pK^{0}_{S}}$ is consistent with the more precise determination from LHCb~\cite{LHCb:2024tnq} and is in agreement with most theoretical calculations.
The decay asymmetry parameters $\alpha_{\Sigma^{+}\pi^{0}}$ and $\alpha_{\Sigma^{0}\pi^{+}}$ agree with the calculation under $SU(3)$ flavor symmetry but $\alpha_{\Sigma^{+}\pi^{0}}$ differs by more than 2 standard deviations from the measurement by the Belle Collaboration~\cite{Belle:2022bsi}.
Because of the cascading decay in $\Lambda_{c}^{+}\to\Lambda\pi^{+},~\Sigma^{0}\pi^{+}$, and $\Sigma^{+}\pi^{0}$, 
we also obtain the value of $\Delta_{BP}$, and use this to derive $\beta_{BP}$. 
Thus, we are able to extract the $S$- and $P$-wave phase difference, $\Delta_S$.
The results of phase differences in this work are consistent with recent theoretical calculations~\cite{Zhong:2024zme,Zhong:2024qqs,Cheng:2024lsn} under the convention $\text{sign}=+1$ as proposed in Ref.~\cite{Wang:2024wrm}.
The extra parameter ``sign'' is introduced in the unified parametrization of $S$- and $P$-wave amplitudes in baryon decays and is important when conducting global analyses of baryon polarization,  involving the strong phase differences, among different experiments and performing future searches for baryon $C\!P$ violation~\cite{Wang:2024wrm}.
We have also derived the dynamic parameters $|A|$ and $|B|$ for weak interactions. 
Previous theoretical studies on the decay of charmed baryons have also calculated $|A|$ and $|B|$, but most of them did not account for the contribution of phase differences to the decay asymmetry parameters, which complicates the accurate calculation of these parameters. 
Only a few recent works have emphasized the impact of phase differences and predicted $|A|$ and $|B|$~\cite{Zhong:2024zme,Zhong:2024qqs,Cheng:2024lsn}. 
Clearly, the reliability of their calculations is improving. 
The results presented here supersede those in Refs.~\cite{BESIII:2019odb, BESIII:2023rwv}.

The phase differences between partial wave amplitudes, $\Delta_S$,  are calculated in the three channels $\Lambda\pi^{+}$, $\Sigma^{0}\pi^{+}$, and $\Sigma^{+}\pi^{0}$, and summarized in Table~\ref{tab:Asycpc}.
The relative phase $\Delta_S$ in channel $\Lambda\pi^{+}$ is consistent with the LHCb measurement~\cite{LHCb:2024tnq}.
Such a phase difference may arise from rescattering processes and loop effects~\cite{Geng:2023pkr,Jia:2024pyb}, suggesting that final-state interactions play an important role in the nonperturbative QCD dynamics in charmed baryon decays.
For the $\Lambda_c^+\to\Sigma^+\pi^0$ channel, the value of $\tan\Delta_S$ differs, with a significance of approximately $3\sigma$, from zero.
This finding is a clear indication of a significant final-state interaction in this channel.
A similar, albeit less significant, effect is seen in $\Lambda_c^+\to\Sigma^0\pi^+$. 
Contrary to hyperon nonleptonic decays, where the excess energy is low, this effect cannot be elastic rescattering but must be understood in terms of strongly coupled channel interactions of the strangeness $S=-1$ and isospin $I=1,2$ systems.

The decays studied in this work are CF processes dominated by tree-level contributions, and therefore, the $C\!P$-violating observables $A_{C\!P}^{\alpha}$ and $\tan\phi_{C\!P}$ are expected to be zero.
The obtained results shown in Table~\ref{tab:Asycpv} align with the SM expectations.
Precision measurements of the decay dynamics in CF modes provide direct access to the underlying nonperturbative QCD effects in the charm sector. 
Furthermore, the information on strong phase differences between different partial waves extracted from CF decays can be used to constrain the corresponding hadronic parameters in CS processes. 
Such constraints are essential inputs for improving theoretical calculations of $C\!P$-violating effects in charmed baryon decays~\cite{He:2024unv,Geng:2023pkr,Zhong:2024qqs}.

In conclusion, the observation of the transverse polarization effect of the $\Lambda_{c}^{+}$ with the BESIII detector represents a complete description in charmed baryon production in $e^+e^-$ annihilations, contributing to our understanding of its production properties. 
Our analysis of decay asymmetry parameters for $\Lambda_{c}^{+}$ decays into $pK^{0}_{S}$, $\Lambda\pi^{+}$, $\Sigma^{0}\pi^{+}$, and $\Sigma^{+}\pi^{0}$ not only updates and validates previous measurements but also reveals significant strong phase differences, providing valuable constraints on various theoretical models. 
Furthermore, the systematic study of production and decay properties provides a comprehensive dataset that refines global fits in flavor physics, enhancing the understanding of QCD in the charm sector and the capability of predicting the $C\!P$ violation in charmed baryons. 
These findings, combined with the precise measurement of polarization dynamics, open new avenues for investigating hadronic interactions, spin effects, and charm quark dynamics. 
This work establishes a robust framework for future experimental programs at Belle II, LHCb, STCF~\cite{Wang:2025yjs} and other facilities, setting the stage for deeper insights into the nonperturbative aspects of QCD and better sensitivities in searching for $C\!P$ violations, as well as the exploration of potential new physics.

\section*{ACKNOWLEDGEMENT}
The BESIII Collaboration thanks the staff of BEPCII (https://cstr.cn/31109.02.BEPC) and the IHEP computing center for their strong support. This work is supported in part by National Key R\&D Program of China under Contracts Nos. 2023YFA1606000, 2023YFA1606704; National Natural Science Foundation of China (NSFC) under Contracts Nos. 11635010, 11935015, 11935016, 11935018, 12025502, 12035009, 12035013, 12061131003, 12192260, 12192261, 12192262, 12192263, 12192264, 12192265, 12221005, 12225509, 12235017, 12361141819, 12422504, 12365015, 125B2106; the Chinese Academy of Sciences (CAS) Large-Scale Scientific Facility Program; CAS under Contract No. YSBR-101; 100 Talents Program of CAS; Fundamental Research Funds for the Central Universities, Lanzhou University under Contracts Nos. lzujbky-2025-ytB01, lzujbky-2023-stlt01, lzujbky-2023-it32, University of Chinese Academy of Sciences; The Natural Science Foundation of Inner Mongolia Autonomous Region No. 2023QN01011; The Institute of Nuclear and Particle Physics (INPAC) and Shanghai Key Laboratory for Particle Physics and Cosmology; ERC under Contract No. 758462; German Research Foundation DFG under Contract No. FOR5327; Istituto Nazionale di Fisica Nucleare, Italy; Knut and Alice Wallenberg Foundation under Contracts Nos. 2021.0174, 2021.0299; Ministry of Development of Turkey under Contract No. DPT2006K-120470; National Research Foundation of Korea under Contract No. NRF-2022R1A2C1092335; National Science and Technology fund of Mongolia; Polish National Science Centre under Contract No. 2024/53/B/ST2/00975; STFC (United Kingdom); Swedish Research Council under Contract No. 2019.04595; U. S. Department of Energy under Contract No. DE-FG02-05ER41374.

\section*{DATA AVAILABILITY}
The data analyzed in this study are not publicly available due to collaboration policies. 
Access to the underlying datasets may be granted upon reasonable request and with permission from the BESIII Collaboration by contacting besiii-publications@ihep.ac.cn.

\appendix
\renewcommand{\thesection}{\Alph{section}}
\renewcommand{\thetable}{\thesection\arabic{table}}
\makeatletter
\@addtoreset{table}{section}
\makeatother
 
\begin{widetext}
\section{CORRELATION MATRIX OF THE FIT PARAMETERS}\label{App:A}
In this appendix, we provide the full correlation matrices shown in Table~\ref{Res:CPCcov} under the baryon $C\!P$-conserving assumption and Table~\ref{Res:CPVcov} without assumption of baryon $C\!P$ conservation. 
These correlations are taken into account in the calculation of the derived observables of $\langle\beta_{BP}\rangle$, $\langle\gamma_{BP}\rangle$, $\langle\Delta_S\rangle$, $A^{\alpha_{BP}}_{C\!P}$, $\tan\phi_{C\!P}$, and $\tan\Delta_S$.
\begin{table}[htbp]
	\small
    \centering
	\caption{Correlation matrix under the assumption of baryon $C\!P$ conservation. Some negligible terms are denoted by ``neg.''}
	\label{Res:CPCcov}
	\renewcommand{\arraystretch}{1.5} 
	\begin{tabular}{l c c c c c c c}
		\hline\hline
		Parameters                 & $\langle\alpha_{pK^{0}_{S}}\rangle$ & $\langle\alpha_{\Lambda\pi^{+}}\rangle$ & $\langle\Delta_{\Lambda\pi^{+}}\rangle$ & $\langle\alpha_{\Sigma^0\pi^{+}}\rangle$ & $\langle\Delta_{\Sigma^0\pi^{+}}\rangle$ & $\langle\alpha_{\Sigma^+\pi^{0}}\rangle$ & $\langle\Delta_{\Sigma^+\pi^{0}}\rangle$ \\
        \hline
        $\langle\alpha_{pK^{0}_{S}}\rangle$      & $ 1.000$              & $ 0.026$                  & $ 0.007$                  & $ 0.014$                   & $ 0.003$                   & $-0.004$                   & $-0.002$ \\
        $\langle\alpha_{\Lambda\pi^{+}}\rangle$  & \raisebox{0.5ex}{...} & $ 1.000$                  & $ 0.002$                  & $ 0.001$                   & $-0.001$                   & $-0.001$                   & $ 0.004$ \\
		$\langle\Delta_{\Lambda\pi^{+}}\rangle$  & \raisebox{0.5ex}{...} & \raisebox{0.5ex}{...}     & $ 1.000$                  & $ 0.009$                   & $-0.001$                   & neg                        & $ 0.014$ \\
        $\langle\alpha_{\Sigma^0\pi^{+}}\rangle$ & \raisebox{0.5ex}{...} & \raisebox{0.5ex}{...}     & \raisebox{0.5ex}{...}     & $ 1.000$                   & $ 0.023$                   & $-0.001$                   & $ 0.001$  \\
		$\langle\Delta_{\Sigma^0\pi^{+}}\rangle$ & \raisebox{0.5ex}{...} & \raisebox{0.5ex}{...}     & \raisebox{0.5ex}{...}     & \raisebox{0.5ex}{...}      & $ 1.000$                   & neg                        & $-0.002$ \\
        $\langle\alpha_{\Sigma^+\pi^{0}}\rangle$ & \raisebox{0.5ex}{...} & \raisebox{0.5ex}{...}     & \raisebox{0.5ex}{...}     & \raisebox{0.5ex}{...}      & \raisebox{0.5ex}{...}      & $ 1.000$                   & $ 0.023$ \\
		$\langle\Delta_{\Sigma^+\pi^{0}}\rangle$ & \raisebox{0.5ex}{...} & \raisebox{0.5ex}{...}     & \raisebox{0.5ex}{...}     & \raisebox{0.5ex}{...}      & \raisebox{0.5ex}{...}      & \raisebox{0.5ex}{...}      & $ 1.000$ \\
		\hline\hline
	\end{tabular}
\end{table}

\begin{table}[H]
	\small
    \centering
	\caption{Correlation matrix without assumption of baryon $C\!P$ conservation. Some negligible terms are denoted by ``neg.''}
	\label{Res:CPVcov}
	\renewcommand{\arraystretch}{1.5} 
	\begin{tabular}{l c c c c c c c c c c c c c c}
		\hline\hline
		Parameters                 & $\alpha_{pK^{0}_{S}}$ & $\bar{\alpha}_{pK^{0}_{S}}$ & $\alpha_{\Lambda\pi^{+}}$ & $\Delta_{\Lambda\pi^{+}}$ & $\bar\alpha_{\Lambda\pi^{+}}$ & $\bar\Delta_{\Lambda\pi^{+}}$ & $\alpha_{\Sigma^0\pi^{+}}$ & $\Delta_{\Sigma^0\pi^{+}}$ & $\bar\alpha_{\Sigma^0\pi^{+}}$ & $\bar\Delta_{\Sigma^0\pi^{+}}$ & $\alpha_{\Sigma^+\pi^{0}}$ & $\Delta_{\Sigma^+\pi^{0}}$ & $\bar\alpha_{\Sigma^+\pi^{0}}$ & $\bar\Delta_{\Sigma^+\pi^{0}}$ \\
        \hline
$\alpha_{pK^{0}_{S}}$ & $1.000$ & $-0.139$ & $0.021$ & $0.005$ & $-0.005$ & $0.001$ & $0.008$ & $0.007$ & $-0.008$ & $-0.007$ & $0.001$ & $0.004$ & $0.003$ & $0.002$\\
$\bar{\alpha}_{pK^{0}_{S}}$ & \raisebox{0.5ex}{...} & $1.000$ & $-0.024$ & $0.002$ & $0.007$ & $0.022$ & neg & $0.005$ & $0.015$ & $0.004$ & $0.001$ & $0.013$ & $-0.009$ & $0.014$\\
$\alpha_{\Lambda\pi^{+}}$ & \raisebox{0.5ex}{...} & \raisebox{0.5ex}{...} & $1.000$ & $-0.003$ & $-0.001$ & $-0.005$ & $-0.001$ & $-0.005$ & neg & $0.002$ & $-0.001$ & $0.001$ & $0.001$ & $-0.003$\\
$\Delta_{\Lambda\pi^{+}}$ & \raisebox{0.5ex}{...} & \raisebox{0.5ex}{...} & \raisebox{0.5ex}{...} & $1.000$ & $0.002$ & $0.026$ & neg & $-0.004$ & neg & $-0.015$ & $-0.002$ & $0.004$ & $-0.002$ & $-0.002$\\
$\bar\alpha_{\Lambda\pi^{+}}$ & \raisebox{0.5ex}{...} & \raisebox{0.5ex}{...} & \raisebox{0.5ex}{...} & \raisebox{0.5ex}{...} & $1.000$ & $-0.007$ & $-0.002$ & $-0.002$ & $0.002$ & $-0.012$ & neg & $-0.002$ & neg & $0.002$\\
$\bar\Delta_{\Lambda\pi^{+}}$ & \raisebox{0.5ex}{...} & \raisebox{0.5ex}{...} & \raisebox{0.5ex}{...} & \raisebox{0.5ex}{...} & \raisebox{0.5ex}{...} & $1.000$ & $-0.002$ & $0.017$ & $0.010$ & $0.003$ & $0.004$ & $0.003$ & $-0.003$ & $0.023$\\
$\alpha_{\Sigma^0\pi^{+}}$ & \raisebox{0.5ex}{...} & \raisebox{0.5ex}{...} & \raisebox{0.5ex}{...} & \raisebox{0.5ex}{...} & \raisebox{0.5ex}{...} & \raisebox{0.5ex}{...} & $1.000$ & $-0.003$ & $0.001$ & $0.003$ & neg & $0.002$ & $-0.002$ & $-0.004$\\
$\Delta_{\Sigma^0\pi^{+}}$ & \raisebox{0.5ex}{...} & \raisebox{0.5ex}{...} & \raisebox{0.5ex}{...} & \raisebox{0.5ex}{...} & \raisebox{0.5ex}{...} & \raisebox{0.5ex}{...} & \raisebox{0.5ex}{...} & $1.000$ & $-0.002$ & $0.004$ & $0.003$ & $0.002$ & $0.001$ & $0.002$\\
$\bar\alpha_{\Sigma^0\pi^{+}}$ & \raisebox{0.5ex}{...} & \raisebox{0.5ex}{...} & \raisebox{0.5ex}{...} & \raisebox{0.5ex}{...} & \raisebox{0.5ex}{...} & \raisebox{0.5ex}{...} & \raisebox{0.5ex}{...} & \raisebox{0.5ex}{...} & $1.000$ & $0.013$ & $0.001$ & $0.004$ & $-0.003$ & $0.006$\\
$\bar\Delta_{\Sigma^0\pi^{+}}$ & \raisebox{0.5ex}{...} & \raisebox{0.5ex}{...} & \raisebox{0.5ex}{...} & \raisebox{0.5ex}{...} & \raisebox{0.5ex}{...} & \raisebox{0.5ex}{...} & \raisebox{0.5ex}{...} & \raisebox{0.5ex}{...} & \raisebox{0.5ex}{...} & $1.000$ & $0.001$ & $0.008$ & $0.001$ & $-0.005$\\
$\alpha_{\Sigma^+\pi^{0}}$ & \raisebox{0.5ex}{...} & \raisebox{0.5ex}{...} & \raisebox{0.5ex}{...} & \raisebox{0.5ex}{...} & \raisebox{0.5ex}{...} & \raisebox{0.5ex}{...} & \raisebox{0.5ex}{...} & \raisebox{0.5ex}{...} & \raisebox{0.5ex}{...} & \raisebox{0.5ex}{...} & $1.000$ & $0.041$ & neg & $0.005$\\
$\Delta_{\Sigma^+\pi^{0}}$ & \raisebox{0.5ex}{...} & \raisebox{0.5ex}{...} & \raisebox{0.5ex}{...} & \raisebox{0.5ex}{...} & \raisebox{0.5ex}{...} & \raisebox{0.5ex}{...} & \raisebox{0.5ex}{...} & \raisebox{0.5ex}{...} & \raisebox{0.5ex}{...} & \raisebox{0.5ex}{...} & \raisebox{0.5ex}{...} & $1.000$ & $-0.001$ & $-0.007$\\
$\bar\alpha_{\Sigma^+\pi^{0}}$ & \raisebox{0.5ex}{...} & \raisebox{0.5ex}{...} & \raisebox{0.5ex}{...} & \raisebox{0.5ex}{...} & \raisebox{0.5ex}{...} & \raisebox{0.5ex}{...} & \raisebox{0.5ex}{...} & \raisebox{0.5ex}{...} & \raisebox{0.5ex}{...} & \raisebox{0.5ex}{...} & \raisebox{0.5ex}{...} & \raisebox{0.5ex}{...} & $1.000$ & $-0.047$\\
$\bar\Delta_{\Sigma^+\pi^{0}}$ & \raisebox{0.5ex}{...} & \raisebox{0.5ex}{...} & \raisebox{0.5ex}{...} & \raisebox{0.5ex}{...} & \raisebox{0.5ex}{...} & \raisebox{0.5ex}{...} & \raisebox{0.5ex}{...} & \raisebox{0.5ex}{...} & \raisebox{0.5ex}{...} & \raisebox{0.5ex}{...} & \raisebox{0.5ex}{...} & \raisebox{0.5ex}{...} & \raisebox{0.5ex}{...} & $1.000$\\
		\hline\hline
	\end{tabular}
\end{table}

\section{RESULTS IN SEPARATE BARYON CHARGES}\label{App:B}
The fitted results of $\alpha_{BP}$, $\Delta_{BP}$, $\bar\alpha_{BP}$, and $\bar\Delta_{BP}$ are summarized in Table~\ref{Res:CPV}, and they are measured without assumption of baryon $C\!P$ conservation.
 \begin{table}[H]
	\centering
	\small
	\caption{Results of $\alpha_{BP}$, $\Delta_{BP}$, $\bar\alpha_{BP}$, and $\bar\Delta_{BP}$. The unit of $\Delta_{BP}$, and $\bar\Delta_{BP}$ is rad, and $\alpha_{BP}$ and $\bar\alpha_{BP}$ are dimensionless numbers.}
	\label{Res:CPV}
	\renewcommand{\arraystretch}{1.5} 
	\begin{tabular}{l c c c c}
		\hline\hline
		Processes                                   & $\alpha_{BP}$                       & $\Delta_{BP}$              & $\bar\alpha_{BP}$                   & $\bar\Delta_{BP}$ \\
		\hline
        $\Lambda_c^+\to pK^{0}_{S}$                 & $-1.000\pm0.187\pm 0.036$           & \raisebox{0.5ex}{...}      & $ 0.824\pm0.169\pm 0.029$           & \raisebox{0.5ex}{...} \\
        $\Lambda_c^+\to\Lambda\pi^{+}$              & $-0.796\pm 0.051\pm 0.009$          & $ 0.343\pm 0.620\pm 0.020$ & $ 0.786\pm 0.056\pm 0.024$          & $-1.166\pm 0.764\pm 0.035$ \\
        $\Lambda_c^+\to\Sigma^{0}\pi^{+}$           & $-0.612\pm 0.109\pm 0.011$          & $ 0.381\pm 1.066\pm 0.042$ & $ 0.403\pm 0.115\pm 0.018$          & $-1.360\pm 0.804\pm 0.031$ \\
        $\Lambda_c^+\to\Sigma^{+}\pi^{0}$           & $-0.537\pm 0.064\pm 0.018$          & $ 2.090\pm 0.786\pm 0.064$ & $ 0.639\pm 0.080\pm 0.016$          & $-1.446\pm 1.315\pm 0.162$ \\
		\hline\hline
	\end{tabular}
\end{table}

\section{Correlation Matrix of the Observed Parameters in Each Process}\label{App:C}
In this appendix, we provide the full correlation matrices of all observed parameters$-$the $\langle\alpha_{BP}\rangle$ and $\langle\Delta_{BP}\rangle$ are fit parameters, and others are derived parameters. 
Only three processes$-$$\Lambda^+_c\to \Lambda\pi^+$, $\Lambda^+_c\to \Sigma^0\pi^+$, and $\Lambda^+_c\to \Sigma^+\pi^0$$-$are provided since the $\Lambda^+_c\to pK^0_S$ process has no derived parameters that could be detected in this work. 
The correlation coefficient are shown in Tables~\ref{Res:CPCcov1}$-$\ref{Res:CPCcov3}. 
The large correlations observed for some derived parameters originate from their common dependence on the same fitted quantities.
\begin{table}[htbp]
	\small
    \centering
	\caption{Correlation matrix of the observed parameters for the $\Lambda^+_c\to \Lambda\pi^+$ decay.}
	\label{Res:CPCcov1}
	\renewcommand{\arraystretch}{1.5} 
	\begin{tabular}{l c c c c c c c}
		\hline\hline
		Parameters                              & $\langle\alpha_{\Lambda\pi^{+}}\rangle$ & $\langle\Delta_{\Lambda\pi^{+}}\rangle$ & $\langle\beta_{\Lambda\pi^{+}}\rangle$ & $\langle\gamma_{\Lambda\pi^{+}}\rangle$ & $\langle\Delta_S\rangle$ & $|A|$ & $|B|$ \\
        \hline
        $\langle\alpha_{\Lambda\pi^{+}}\rangle$               & $ 1.000$                  & $ 0.001$                  & $ 0.111$                 & $ 0.188$                  & $-0.152$                               & $ 0.177$               & $-0.211$ \\
		$\langle\Delta_{\Lambda\pi^{+}}\rangle$               & \raisebox{0.5ex}{...}     & $ 1.000$                  & $ 0.963$                 & $-0.899$                  & $-0.941$                               & $-0.892$               & $ 0.907$ \\
        $\langle\beta_{\Lambda\pi^{+}}\rangle$                & \raisebox{0.5ex}{...}     & \raisebox{0.5ex}{...}     & $ 1.000$                 & $-0.768$                  & $-0.996$                               & $-0.753$               & $ 0.791$ \\
		$\langle\gamma_{\Lambda\pi^{+}}\rangle$               & \raisebox{0.5ex}{...}     & \raisebox{0.5ex}{...}     & \raisebox{0.5ex}{...}    & $ 1.000$                  & $ 0.717$                               & $ 0.999$               & $-0.996$ \\
        $\langle\Delta_S\rangle$                              & \raisebox{0.5ex}{...}     & \raisebox{0.5ex}{...}     & \raisebox{0.5ex}{...}    & \raisebox{0.5ex}{...}     & $ 1.000$                               & $ 0.701$               & $-0.741$ \\
		$|A|_{\Lambda\pi^{+}}$                                & \raisebox{0.5ex}{...}     & \raisebox{0.5ex}{...}     & \raisebox{0.5ex}{...}    & \raisebox{0.5ex}{...}     & \raisebox{0.5ex}{...}                  & $ 1.000$               & $-0.991$ \\
        $|B|_{\Lambda\pi^{+}}$                                & \raisebox{0.5ex}{...}     & \raisebox{0.5ex}{...}     & \raisebox{0.5ex}{...}    & \raisebox{0.5ex}{...}     & \raisebox{0.5ex}{...}                  & \raisebox{0.5ex}{...}  & $ 1.000$ \\
		\hline\hline
	\end{tabular}
\end{table}

\begin{table}[htbp]
	\small
    \centering
	\caption{Correlation matrix of the observed parameters for the $\Lambda^+_c\to \Sigma^0\pi^+$ decay.}
	\label{Res:CPCcov2}
	\renewcommand{\arraystretch}{1.5} 
	\begin{tabular}{l c c c c c c c}
		\hline\hline
		Parameters                               & $\langle\alpha_{\Sigma^0\pi^{+}}\rangle$ & $\langle\Delta_{\Sigma^0\pi^{+}}\rangle$ & $\langle\beta_{\Sigma^0\pi^{+}}\rangle$ & $\langle\gamma_{{\Sigma^0\pi^{+}}}\rangle$ & $\langle\Delta_S\rangle$ & $|A|$ & $|B|$ \\
        \hline
        $\langle\alpha_{\Sigma^0\pi^{+}}\rangle$               & $ 1.000$                   & $ 0.022$                   & $ 0.135$                  & $ 0.023$                     & $-0.265$                                & $ 0.009$                & $-0.050$ \\
		$\langle\Delta_{\Sigma^0\pi^{+}}\rangle$               & \raisebox{0.5ex}{...}      & $ 1.000$                   & $ 0.751$                  & $-0.957$                     & $-0.690$                                & $-0.940$                & $ 0.965$ \\
        $\langle\beta_{\Sigma^0\pi^{+}}\rangle$                & \raisebox{0.5ex}{...}      & \raisebox{0.5ex}{...}      & $ 1.000$                  & $ 0.588$                     & $-0.962$                                & $-0.509$                & $ 0.686$ \\
		$\langle\gamma_{\Sigma^0\pi^{+}}\rangle$               & \raisebox{0.5ex}{...}      & \raisebox{0.5ex}{...}      & \raisebox{0.5ex}{...}     & $ 1.000$                     & $ 0.491$                                & $ 0.992$                & $-0.986$ \\
        $\langle\Delta_S\rangle$                               & \raisebox{0.5ex}{...}      & \raisebox{0.5ex}{...}      & \raisebox{0.5ex}{...}     & \raisebox{0.5ex}{...}        & $ 1.000$                                & $ 0.429$                & $-0.577$ \\
		$|A|_{\Sigma^0\pi^{+}}$                                & \raisebox{0.5ex}{...}      & \raisebox{0.5ex}{...}      & \raisebox{0.5ex}{...}     & \raisebox{0.5ex}{...}        & \raisebox{0.5ex}{...}                   & $ 1.000$                & $-0.960$ \\
        $|B|_{\Sigma^0\pi^{+}}$                                & \raisebox{0.5ex}{...}      & \raisebox{0.5ex}{...}      & \raisebox{0.5ex}{...}     & \raisebox{0.5ex}{...}        & \raisebox{0.5ex}{...}                   & \raisebox{0.5ex}{...}   & $ 1.000$ \\
		\hline\hline
	\end{tabular}
\end{table}

\begin{table}[htbp]
	\small
    \centering
	\caption{Correlation matrix of the observed parameters for the $\Lambda^+_c\to \Sigma^+\pi^0$ decay.}
	\label{Res:CPCcov3}
	\renewcommand{\arraystretch}{1.5} 
	\begin{tabular}{l c c c c c c c}
		\hline\hline
		Parameters                               & $\langle\alpha_{\Sigma^+\pi^{0}}\rangle$ & $\langle\Delta_{\Sigma^+\pi^{0}}\rangle$ & $\langle\beta_{\Sigma^+\pi^{0}}\rangle$ & $\langle\gamma_{\Sigma^+\pi^{0}}\rangle$ & $\langle\Delta_S\rangle$ & $|A|$ & $|B|$ \\
        \hline
        $\langle\alpha_{\Sigma^+\pi^{0}}\rangle$               & $ 1.000$                   & $ 0.022$                   & $ 0.120$                  & $-0.045$                   & $-0.234$                                & $-0.062$                & $ 0.035$ \\
		$\langle\Delta_{\Sigma^+\pi^{0}}\rangle$               & \raisebox{0.5ex}{...}      & $ 1.000$                   & $-0.609$                  & $-0.978$                   & $ 0.573$                                & $-0.983$                & $ 0.966$ \\
        $\langle\beta_{\Sigma^+\pi^{0}}\rangle$                & \raisebox{0.5ex}{...}      & \raisebox{0.5ex}{...}      & $ 1.000$                  & $ 0.497$                   & $-0.974$                                & $ 0.600$                & $-0.410$ \\
		$\langle\gamma_{\Sigma^+\pi^{0}}\rangle$               & \raisebox{0.5ex}{...}      & \raisebox{0.5ex}{...}      & \raisebox{0.5ex}{...}     & $ 1.000$                   & $-0.441$                                & $ 0.990$                & $-0.993$ \\
        $\langle\Delta_S\rangle$                               & \raisebox{0.5ex}{...}      & \raisebox{0.5ex}{...}      & \raisebox{0.5ex}{...}     & \raisebox{0.5ex}{...}      & $ 1.000$                                & $-0.535$                & $ 0.366$ \\
		$|A|_{\Sigma^+\pi^{0}}$                                & \raisebox{0.5ex}{...}      & \raisebox{0.5ex}{...}      & \raisebox{0.5ex}{...}     & \raisebox{0.5ex}{...}      & \raisebox{0.5ex}{...}                   & $ 1.000$                & $-0.969$ \\
        $|B|_{\Sigma^+\pi^{0}}$                                & \raisebox{0.5ex}{...}      & \raisebox{0.5ex}{...}      & \raisebox{0.5ex}{...}     & \raisebox{0.5ex}{...}      & \raisebox{0.5ex}{...}                   & \raisebox{0.5ex}{...}   & $ 1.000$ \\
		\hline\hline
	\end{tabular}
\end{table}

\end{widetext}
\clearpage
\bibliography{bibliography.bib}

\onecolumngrid
\vspace{1em}
\noindent\makebox[\linewidth]{\rule{0.48\linewidth}{0.6pt}}
\vspace{1em}

\begin{center}
    \small
    \input{authorlist_2025-04-10_orcid}
\end{center}

\end{document}

%% file: authorlist_2025-04-10_orcid.tex
M.~Ablikim$^{1}$\BESIIIorcid{0000-0002-3935-619X},
M.~N.~Achasov$^{4,b}$\BESIIIorcid{0000-0002-9400-8622},
P.~Adlarson$^{77}$\BESIIIorcid{0000-0001-6280-3851},
X.~C.~Ai$^{82}$\BESIIIorcid{0000-0003-3856-2415},
R.~Aliberti$^{36}$\BESIIIorcid{0000-0003-3500-4012},
A.~Amoroso$^{76A,76C}$\BESIIIorcid{0000-0002-3095-8610},
Q.~An$^{73,59,\dagger}$,
Y.~Bai$^{58}$\BESIIIorcid{0000-0001-6593-5665},
O.~Bakina$^{37}$\BESIIIorcid{0009-0005-0719-7461},
Y.~Ban$^{47,g}$\BESIIIorcid{0000-0002-1912-0374}
% ...
\\
\vspace{0.2cm}
(BESIII Collaboration)\\
\vspace{0.2cm} {\it
$^{1}$ Institute of High Energy Physics, Beijing 100049, People's Republic of China\\
% ...
$^{4}$ Budker Institute of Nuclear Physics SB RAS (BINP), Novosibirsk 630090, Russia\\
% ...
$^{36}$ Johannes Gutenberg University of Mainz, Johann-Joachim-Becher-Weg 45, D-55099 Mainz, Germany\\
$^{37}$ Joint Institute for Nuclear Research, 141980 Dubna, Moscow region, Russia\\
% ...
$^{47}$ Peking University, Beijing 100871, People's Republic of China\\
% ...
$^{58}$ Southeast University, Nanjing 211100, People's Republic of China\\
$^{59}$ State Key Laboratory of Particle Detection and Electronics, Beijing 100049, Hefei 230026, People's Republic of China\\
% ...
$^{73}$ University of Science and Technology of China, Hefei 230026, People's Republic of China\\
% ...
$^{76}$ University of Turin and INFN, (A)University of Turin, I-10125, Turin, Italy; (B)University of Eastern Piedmont, I-15121, Alessandria, Italy; (C)INFN, I-10125, Turin, Italy\\
$^{77}$ Uppsala University, Box 516, SE-75120 Uppsala, Sweden\\
% ...
$^{82}$ Zhengzhou University, Zhengzhou 450001, People's Republic of China\\

\vspace{0.2cm}
$^{\dagger}$ Deceased\\
% ...
$^{b}$ Also at the Novosibirsk State University, Novosibirsk, 630090, Russia\\
% ...
$^{g}$ Also at State Key Laboratory of Nuclear Physics and Technology, Peking University, Beijing 100871, People's Republic of China\\
% ...

}

%% file: bibliography.bib
@article{LHCb:2019hro,
    author = "Aaij, Roel and others",
    collaboration = "LHCb Collaboration",
    title = "{Observation of $C\!P$ Violation in Charm Decays}",
    eprint = "1903.08726",
    archivePrefix = "arXiv",
    primaryClass = "hep-ex",
    reportNumber = "LHCb-PAPER-2019-006, CERN-EP-2019-042",
    doi = "10.1103/PhysRevLett.122.211803",
    journal = "Phys. Rev. Lett.",
    volume = "122",
    number = "21",
    pages = "211803",
    year = "2019"
}

@article{BESIII:2018cnd,
    author = "Ablikim, M. and others",
    collaboration = "BESIII Collaboration",
    title = "{Polarization and Entanglement in Baryon-Antibaryon Pair Production in Electron-Positron Annihilation}",
    eprint = "1808.08917",
    archivePrefix = "arXiv",
    primaryClass = "hep-ex",
    doi = "10.1038/s41567-019-0494-8",
    journal = "Nature Phys.",
    volume = "15",
    pages = "631--634",
    year = "2019"
}

@article{BESIII:2022qax,
    author = "Ablikim, M. and others",
    collaboration = "BESIII Collaboration",
    title = "{Precise Measurements of Decay Parameters and $C\!P$ Asymmetry with Entangled $\Lambda-\bar{\Lambda}$ Pairs Pairs}",
    eprint = "2204.11058",
    archivePrefix = "arXiv",
    primaryClass = "hep-ex",
    doi = "10.1103/PhysRevLett.129.131801",
    journal = "Phys. Rev. Lett.",
    volume = "129",
    number = "13",
    pages = "131801",
    year = "2022"
}

@article{BESIII:2023jhj,
    author = "Ablikim, M. and others",
    collaboration = "BESIII Collaboration",
    title = "{Investigation of the $\Delta I = 1/2$ Rule and Test of $C\!P$ Symmetry through the Measurement of Decay Asymmetry Parameters in $\Xi^-$ Decays}",
    eprint = "2309.14667",
    archivePrefix = "arXiv",
    primaryClass = "hep-ex",
    doi = "10.1103/PhysRevLett.132.101801",
    journal = "Phys. Rev. Lett.",
    volume = "132",
    number = "10",
    pages = "101801",
    year = "2024"
}

@article{BESIII:2023drj,
    author = "Ablikim, Medina and others",
    collaboration = "BESIII Collaboration",
    title = "{Tests of $C\!P$ symmetry in entangled $\Xi^0$-$\bar{\Xi}^0$ pairs}",
    eprint = "2305.09218",
    archivePrefix = "arXiv",
    primaryClass = "hep-ex",
    doi = "10.1103/PhysRevD.108.L031106",
    journal = "Phys. Rev. D",
    volume = "108",
    number = "3",
    pages = "L031106",
    year = "2023"
}

@article{BESIII:2021ypr,
    author = "Ablikim, Medina and others",
    collaboration = "BESIII Collaboration",
    title = "{Probing $C\!P$ symmetry and weak phases with entangled double-strange baryons}",
    eprint = "2105.11155",
    archivePrefix = "arXiv",
    primaryClass = "hep-ex",
    doi = "10.1038/s41586-022-04624-1",
    journal = "Nature",
    volume = "606",
    number = "7912",
    pages = "64--69",
    year = "2022"
}

@article{BESIII:2022lsz,
    author = "Ablikim, Medina and others",
    collaboration = "BESIII Collaboration",
    title = "{Observation of $\Xi^-$ hyperon transverse polarization in \ensuremath{\psi}(3686)\textrightarrow{}$\Xi^-$$\bar{\Xi}^+$}",
    eprint = "2206.10900",
    archivePrefix = "arXiv",
    primaryClass = "hep-ex",
    doi = "10.1103/PhysRevD.106.L091101",
    journal = "Phys. Rev. D",
    volume = "106",
    number = "9",
    pages = "L091101",
    year = "2022"
}

@article{BESIII:2023rwv,
    author = "Ablikim, Medina and others",
    collaboration = "BESIII Collaboration",
    title = "{Measurement of Energy-Dependent Pair-Production Cross Section and Electromagnetic Form Factors of a Charmed Baryon}",
    eprint = "2307.07316",
    archivePrefix = "arXiv",
    primaryClass = "hep-ex",
    doi = "10.1103/PhysRevLett.131.191901",
    journal = "Phys. Rev. Lett.",
    volume = "131",
    number = "19",
    pages = "191901",
    year = "2023"
}

@article{BESIII:2009fln,
    author = "Ablikim, M. and others",
    collaboration = "BESIII Collaboration",
    title = "{Design and Construction of the BESIII Detector}",
    eprint = "0911.4960",
    archivePrefix = "arXiv",
    primaryClass = "physics.ins-det",
    doi = "10.1016/j.nima.2009.12.050",
    journal = "Nucl. Instrum. Meth. A",
    volume = "614",
    pages = "345--399",
    year = "2010"
}

@inproceedings{Yu:2016cof,
    author = "Yu, Chenghui and others",
    title = "{BEPCII Performance and Beam Dynamics Studies on Luminosity}",
    booktitle = "{7th International Particle Accelerator Conference}",
    doi = "10.18429/JACoW-IPAC2016-TUYA01",
    pages = "TUYA01",
    year = "2016"
}

@article{BESIII:2020nme,
    author = "Ablikim, M. and others",
    collaboration = "BESIII Collaboration",
    title = "{Future Physics Programme of BESIII}",
    eprint = "1912.05983",
    archivePrefix = "arXiv",
    primaryClass = "hep-ex",
    reportNumber = "HEP-Physics-Report-BESIII-2019-12-13",
    doi = "10.1088/1674-1137/44/4/040001",
    journal = "Chin. Phys. C",
    volume = "44",
    number = "4",
    pages = "040001",
    year = "2020"
}

@article{Lu2020-hf,
  title    = "Online monitoring of the center-of-mass energy from real data at {BESIII}",
  author   = "Lu, Jiada and Xiao, Yanjia and Ji, Xiaobin",
  doi = "10.1007/s41605-020-00188-8",
  journal  = "Radiat. Detect. Technol. Meth.",
  volume   =  4,
  number   =  3,
  pages    = "337--344",
  year     =  2020
}

@article{Zhang2022-id,
  title    = "Suppression of top-up injection backgrounds with offline event filter in the {BESIII} experiment",
  author   = "Zhang, Jia-Wei and others",
  doi = "10.1007/s41605-022-00331-7",
  journal  = "Radiat. Detect. Technol. Meth.",
  volume   =  6,
  number   =  3,
  pages    = "289--293",
  year     =  2022
}

@article{Guo2017-of,
  title    = "The study of time calibration for upgraded end cap TOF of {BESIII}",
  author   = "Guo, Ying-Xiao and others",
  doi = "10.1007/s41605-017-0012-4",
  journal  = "Radiat. Detect. Technol. Meth.",
  volume   =  1,
  number   =  15,
  year     =  2017
}

@article{Li2017-zt,
  title    = "Study of {MRPC} technology for {BESIII} endcap-TOF upgrade",
  author   = "Li, Xin and others",
  doi = "10.1007/s41605-017-0014-2",
  journal  = "Radiat. Detect. Technol. Meth.",
  volume   =  1,
  number   =  13,
  year     =  2017
}

@article{Cao2020-vl,
  title    = "Design and construction of the new {BESIII} endcap Time-of-Flight system with {MRPC} Technology",
  author   = "Cao, P, Chen and others",
  doi = "10.1016/j.nima.2019.163053",
  journal  = "Nucl. Instrum. Methods Phys. Res. A",
  volume   =  953,
  number   =  163053,
  pages    = "163053",
  year     =  2020
}

@article{BESIII:2022dxl,
    author = "Ablikim, M. and others",
    collaboration = "BESIII Collaboration",
    title = "{Measurement of integrated luminosities at BESIII for data samples at center-of-mass energies between 4.0 and 4.6 GeV}",
    eprint = "2203.03133",
    archivePrefix = "arXiv",
    primaryClass = "hep-ex",
    doi = "10.1088/1674-1137/ac80b4",
    journal = "Chin. Phys. C",
    volume = "46",
    number = "11",
    pages = "113002",
    year = "2022"
}

@article{BESIII:2022ulv,
    author = "Ablikim, M. and others",
    collaboration = "BESIII Collaboration",
    title = "{Luminosities and energies of $e^{+}e^{-}$ collision data taken between 4.61 GeV and 4.95 GeV at BESIII}",
    eprint = "2205.04809",
    archivePrefix = "arXiv",
    primaryClass = "hep-ex",
    doi = "10.1088/1674-1137/ac84cc",
    journal = "Chin. Phys. C",
    volume = "46",
    number = "11",
    pages = "113003",
    year = "2022"
}

@article{GEANT4:2002zbu,
    author = "Agostinelli, S. and others",
    collaboration = "GEANT4 Collaboration",
    title = "{GEANT4--a simulation toolkit}",
    reportNumber = "SLAC-PUB-9350, FERMILAB-PUB-03-339, CERN-IT-2002-003",
    doi = "10.1016/S0168-9002(03)01368-8",
    journal = "Nucl. Instrum. Meth. A",
    volume = "506",
    pages = "250--303",
    year = "2003"
}

@article{Jadach:2000ir,
    author = "Jadach, S. and Ward, B. F. L. and Was, Z.",
    title = "{Coherent exclusive exponentiation for precision Monte Carlo calculations}",
    eprint = "hep-ph/0006359",
    archivePrefix = "arXiv",
    reportNumber = "CERN-TH-2000-087, UTHEP-99-09-01",
    doi = "10.1103/PhysRevD.63.113009",
    journal = "Phys. Rev. D",
    volume = "63",
    pages = "113009",
    year = "2001"
}

@article{Jadach:1999vf,
    author = "Jadach, S. and Ward, B. F. L. and Was, Z.",
    title = "{The Precision Monte Carlo event generator $KK$ for two fermion final states in $e^{+}e^{-}$ collisions}",
    eprint = "hep-ph/9912214",
    archivePrefix = "arXiv",
    reportNumber = "DESY-99-106, CERN-TH-99-235, UTHEP-99-08-01",
    doi = "10.1016/S0010-4655(00)00048-5",
    journal = "Comput. Phys. Commun.",
    volume = "130",
    pages = "260--325",
    year = "2000"
}

@article{Lange:2001uf,
    author = "Lange, D. J.",
    editor = "Erhan, S. and Schlein, P. and Rozen, Y.",
    title = "{The EvtGen particle decay simulation package}",
    doi = "10.1016/S0168-9002(01)00089-4",
    journal = "Nucl. Instrum. Meth. A",
    volume = "462",
    pages = "152--155",
    year = "2001"
}

@article{Ping:2008zz,
    author = "Ping, Rong-Gang",
    title = "{Event generators at {BESIII}}",
    doi = "10.1088/1674-1137/32/8/001",
    journal = "Chin. Phys. C",
    volume = "32",
    pages = "599",
    year = "2008"
}

@article{ParticleDataGroup:2024cfk,
    author = "Navas, S. and others",
    collaboration = "Particle Data Group",
    title = "{Review of particle physics}",
    doi = "10.1103/PhysRevD.110.030001",
    journal = "Phys. Rev. D",
    volume = "110",
    number = "3",
    pages = "030001",
    year = "2024"
}

@article{Chen:2000tv,
    author = "Chen, J. C. and Huang, G. S. and Qi, X. R. and Zhang, D. H. and Zhu, Y. S.",
    title = "{Event generator for $J/\psi$ and $\psi(2S)$ decay}",
    doi = "10.1103/PhysRevD.62.034003",
    journal = "Phys. Rev. D",
    volume = "62",
    pages = "034003",
    year = "2000"
}

@article{Yang:2014vra,
    author = "Yang, Rui-Ling and Ping, Rong-Gang and Chen, Hong",
    title = "{Tuning and Validation of the Lundcharm Model with $J/\psi$ Decays}",
    doi = "10.1088/0256-307X/31/6/061301",
    journal = "Chin. Phys. Lett.",
    volume = "31",
    pages = "061301",
    year = "2014"
}

@article{Barberio:1990ms,
    author = "Barberio, Elisabetta and van Eijk, Bob and Was, Zbigniew",
    title = "{PHOTOS: A Universal Monte Carlo for QED radiative corrections in decays}",
    reportNumber = "CERN-TH-5857-90",
    doi = "10.1016/0010-4655(91)90012-A",
    journal = "Comput. Phys. Commun.",
    volume = "66",
    pages = "115--128",
    year = "1991"
}

@article{BESIII:2022xne,
    author = "Ablikim, Medina and others",
    collaboration = "BESIII Collaboration",
    title = "{Observations of the Cabibbo-Suppressed decays $\Lambda_{c}^{+}\to n\pi^{+}\pi^{0}$, $n\pi^{+}\pi^{-}\pi^{+}$ and the Cabibbo-Favored decay $\Lambda_{c}^{+}\to nK^{-}\pi^{+}\pi^{+}$}",
    eprint = "2210.03375",
    archivePrefix = "arXiv",
    primaryClass = "hep-ex",
    doi = "10.1088/1674-1137/ac9d29",
    journal = "Chin. Phys. C",
    volume = "47",
    number = "2",
    pages = "023001",
    year = "2023"
}

@article{LHCb:2021chn,
    author = "Aaij, Roel and others",
    collaboration = "LHCb Collaboration",
    title = "{Evidence for a new structure in the $J/\psi p$ and $J/\psi \bar{p}$ systems in $B_s^0 \to J/\psi p \bar{p}$ decays}",
    eprint = "2108.04720",
    archivePrefix = "arXiv",
    primaryClass = "hep-ex",
    reportNumber = "LHCb-PAPER-2021-018, CERN-EP-2021-150",
    doi = "10.1103/PhysRevLett.128.062001",
    journal = "Phys. Rev. Lett.",
    volume = "128",
    number = "6",
    pages = "062001",
    year = "2022"
}

@article{LHCb:2022ogu,
    author = "Aaij, R. and others",
    collaboration = "LHCb Collaboration",
    title = "{Observation of a $J/\psi \Lambda$ Resonance Consistent with a Strange Pentaquark Candidate in $B^- \to J/\psi \Lambda \bar{p}$ Decays}",
    eprint = "2210.10346",
    archivePrefix = "arXiv",
    primaryClass = "hep-ex",
    reportNumber = "CERN-EP-2022-198, LHCb-PAPER-2022-031",
    doi = "10.1103/PhysRevLett.131.031901",
    journal = "Phys. Rev. Lett.",
    volume = "131",
    number = "3",
    pages = "031901",
    year = "2023"
}

@article{ARGUS:1990hfq,
    author = "Albrecht, H. and others",
    collaboration = "ARGUS Collaboration",
    title = "{Search for Hadronic $b \to u$ Decays}",
    reportNumber = "DESY-90-008",
    doi = "10.1016/0370-2693(90)91293-K",
    journal = "Phys. Lett. B",
    volume = "241",
    pages = "278--282",
    year = "1990"
}

@article{BESIII:2017kqg,
    author = "Ablikim, Medina and others",
    collaboration = "BESIII Collaboration",
    title = "{Precision measurement of the $e^{+}e^{-}~\rightarrow~\Lambda_{c}^{+} \bar{\Lambda}_{c}^{-}$ cross section near threshold}",
    eprint = "1710.00150",
    archivePrefix = "arXiv",
    primaryClass = "hep-ex",
    doi = "10.1103/PhysRevLett.120.132001",
    journal = "Phys. Rev. Lett.",
    volume = "120",
    number = "13",
    pages = "132001",
    year = "2018"
}

@article{LHCb:2022sck,
    author = "Aaij, Roel and others",
    collaboration = "LHCb Collaboration",
    title = "{Amplitude analysis of the $ \Lambda_c^+ \to p K^- \pi^+ $ decay and $ {\Lambda}_c^{+} $ baryon polarization measurement in semileptonic beauty hadron decays}",
    eprint = "2208.03262",
    archivePrefix = "arXiv",
    primaryClass = "hep-ex",
    reportNumber = "LHCb-PAPER-2022-002, CERN-EP-2022-124",
    doi = "10.1103/PhysRevD.108.012023",
    journal = "Phys. Rev. D",
    volume = "108",
    number = "1",
    pages = "012023",
    year = "2023"
}

@article{BESIII:2019odb,
    author = "Ablikim, Medina and others",
    collaboration = "BESIII Collaboration",
    title = "{Measurements of Weak Decay Asymmetries of $\Lambda_c^+\to pK_S^0$, $\Lambda\pi^+$, $\Sigma^+\pi^0$, and $\Sigma^0\pi^+$}",
    eprint = "1905.04707",
    archivePrefix = "arXiv",
    primaryClass = "hep-ex",
    doi = "10.1103/PhysRevD.100.072004",
    journal = "Phys. Rev. D",
    volume = "100",
    number = "7",
    pages = "072004",
    year = "2019"
}

@article{James:1975dr,
    author = "James, F. and Roos, M.",
    title = "{Minuit: A System for Function Minimization and Analysis of the Parameter Errors and Correlations}",
    reportNumber = "CERN-DD-75-20",
    doi = "10.1016/0010-4655(75)90039-9",
    journal = "Comput. Phys. Commun.",
    volume = "10",
    pages = "343--367",
    year = "1975"
}

@article{BESIII:2018ciw,
    author = "Ablikim, Medina and others",
    collaboration = "BESIII Collaboration",
    title = "{Measurement of absolute branching fraction of the inclusive decay $\Lambda_{c}^{+} \to \Lambda + X$}",
    eprint = "1803.05706",
    archivePrefix = "arXiv",
    primaryClass = "hep-ex",
    doi = "10.1103/PhysRevLett.121.062003",
    journal = "Phys. Rev. Lett.",
    volume = "121",
    number = "6",
    pages = "062003",
    year = "2018"
}

@article{LHCb:2023crj,
    author = "Aaij, Roel and others",
    collaboration = "LHCb Collaboration",
    title = "{$ {\Lambda}_c^{+} $ polarimetry using the dominant hadronic mode}",
    eprint = "2301.07010",
    archivePrefix = "arXiv",
    primaryClass = "hep-ex",
    reportNumber = "LHCb-PAPER-2022-044, CERN-EP-2022-287",
    doi = "10.1007/JHEP07(2023)228",
    journal = "JHEP",
    volume = "07",
    pages = "228",
    year = "2023"
}

@article{Lee:1957qs,
    author = "Lee, T. D. and Yang, Chen-Ning",
    title = "{General Partial Wave Analysis of the Decay of a Hyperon of Spin 1/2}",
    doi = "10.1103/PhysRev.108.1645",
    journal = "Phys. Rev.",
    volume = "108",
    pages = "1645--1647",
    year = "1957"
}

@article{Chen:2019hqi,
    author = "Chen, Hong and Ping, Rong-Gang",
    title = "{$\Lambda^+_c$ transverse polarization and decay asymmetry parameters}",
    doi = "10.1103/PhysRevD.99.114027",
    journal = "Phys. Rev. D",
    volume = "99",
    number = "11",
    pages = "114027",
    year = "2019"
}

@article{Belle:2022uod,
    author = "Li, L. K. and others",
    collaboration = "Belle Collaboration",
    title = "{Search for $C\!P$ violation and measurement of branching fractions and decay asymmetry parameters for $\Lambda_{c}^{+}\to\Lambda h^{+}$ and $\Lambda_{c}^{+}\to\Sigma^{0}h^{+} (h=K,\pi)$}",
    eprint = "2208.08695",
    archivePrefix = "arXiv",
    primaryClass = "hep-ex",
    reportNumber = "Belle Preprint 2022-17, KEK Preprint 2022-22, UCHEP-22-04",
    doi = "10.1016/j.scib.2023.02.017",
    journal = "Sci. Bull.",
    volume = "68",
    pages = "583--592",
    year = "2023"
}

@article{Belle:2022bsi,
    author = "Li, S. X. and others",
    collaboration = "Belle Collaboration",
    title = "{Measurements of branching fractions of $\Lambda_c^+ \to \Sigma^+ \eta$ and $\Lambda_c^+ \to \Sigma^+ \eta'$ and asymmetry parameters of $\Lambda_c^+ \to \Sigma^+ \pi^0$, $\Lambda_c^+ \to \Sigma^+ \eta$, and $\Lambda_c^+ \to \Sigma^+ \eta'$}",
    eprint = "2208.10825",
    archivePrefix = "arXiv",
    primaryClass = "hep-ex",
    reportNumber = "Belle Preprint 2022-19; KEK Preprint 2022-25",
    doi = "10.1103/PhysRevD.107.032003",
    journal = "Phys. Rev. D",
    volume = "107",
    pages = "032003",
    year = "2023"
}

@article{BESIII:2023euh,
    author = "Ablikim, Medina and others",
    collaboration = "BESIII Collaboration",
    title = "{Measurement of $\Lambda$ transverse polarization in $e^{+}e^{-}$ collisions at $\sqrt{s}$ = 3.68 -3.71 GeV}",
    eprint = "2303.00271",
    archivePrefix = "arXiv",
    primaryClass = "hep-ex",
    doi = "10.1007/JHEP10(2023)081",
    journal = "JHEP",
    volume = "10",
    pages = "081",
    year = "2023",
    note = "[Erratum: JHEP 12, 080 (2023)]"
}

@article{BESIII:2023ynq,
    author = "Ablikim, M. and others",
    collaboration = "BESIII Collaboration",
    title = "{Determination of the $\Sigma^+$ Timelike Electromagnetic Form Factors}",
    eprint = "2307.15894",
    archivePrefix = "arXiv",
    primaryClass = "hep-ex",
    doi = "10.1103/PhysRevLett.132.081904",
    journal = "Phys. Rev. Lett.",
    volume = "132",
    number = "8",
    pages = "081904",
    year = "2024"
}

@article{BESIII:2024nif,
    author = "Ablikim, M. and others",
    collaboration = "BESIII Collaboration",
    title = "{Strong and Weak $C\!P$ Tests in Sequential Decays of Polarized $\Sigma^0$ Hyperons}",
    eprint = "2406.06118",
    archivePrefix = "arXiv",
    primaryClass = "hep-ex",
    doi = "10.1103/PhysRevLett.133.101902",
    journal = "Phys. Rev. Lett.",
    volume = "133",
    number = "10",
    pages = "101902",
    year = "2024"
}

@article{Perotti:2018wxm,
    author = {Perotti, Elisabetta and F\"aldt, G\"oran and Kupsc, Andrzej and Leupold, Stefan and Song, Jiao Jiao},
    title = "{Polarization observables in $e^+e^-$ annihilation to a baryon-antibaryon pair}",
    eprint = "1809.04038",
    archivePrefix = "arXiv",
    primaryClass = "hep-ph",
    doi = "10.1103/PhysRevD.99.056008",
    journal = "Phys. Rev. D",
    volume = "99",
    number = "5",
    pages = "056008",
    year = "2019"
}

@article{Liu:2023dvg,
    author = "Liu, Chia-Wei",
    title = "{Nonleptonic two-body weak decays of charmed baryons}",
    eprint = "2308.07754",
    archivePrefix = "arXiv",
    primaryClass = "hep-ph",
    doi = "10.1103/PhysRevD.109.033004",
    journal = "Phys. Rev. D",
    volume = "109",
    number = "3",
    pages = "033004",
    year = "2024"
}

@article{Zhong:2022exp,
    author = "Zhong, Huiling and Xu, Fanrong and Wen, Qiaoyi and Gu, Yu",
    title = "{Weak decays of antitriplet charmed baryons from the perspective of flavor symmetry}",
    eprint = "2210.12728",
    archivePrefix = "arXiv",
    primaryClass = "hep-ph",
    doi = "10.1007/JHEP02(2023)235",
    journal = "JHEP",
    volume = "02",
    pages = "235",
    year = "2023"
}

@article{Zou:2019kzq,
    author = "Zou, Jinqi and Xu, Fanrong and Meng, Guanbao and Cheng, Hai-Yang",
    title = "{Two-body hadronic weak decays of antitriplet charmed baryons}",
    eprint = "1910.13626",
    archivePrefix = "arXiv",
    primaryClass = "hep-ph",
    doi = "10.1103/PhysRevD.101.014011",
    journal = "Phys. Rev. D",
    volume = "101",
    number = "1",
    pages = "014011",
    year = "2020"
}

@article{Geng:2023pkr,
    author = "Geng, Chao-Qiang and He, Xiao-Gang and Jin, Xiang-Nan and Liu, Chia-Wei and Yang, Chang",
    title = "{Complete determination of $SU(3)_F$ amplitudes and strong phase in $\Lambda_c^+ \to \Xi^0 K^+$}",
    eprint = "2310.05491",
    archivePrefix = "arXiv",
    primaryClass = "hep-ph",
    doi = "10.1103/PhysRevD.109.L071302",
    journal = "Phys. Rev. D",
    volume = "109",
    number = "7",
    pages = "L071302",
    year = "2024"
}

@MISC{Zhong:2024zme,
    author = "Zhong, Huiling and Xu, Fanrong and Cheng, Hai-Yang",
    title = "{Topological Diagrams and Hadronic Weak Decays of Charmed Baryons}",
    howpublished = "e-print arXiv:2401.15926",
    eprint = "2401.15926",
    archivePrefix = "arXiv",
    primaryClass = "hep-ph",
    month = "1",
    year = "2024"
}

@article{Chen:2023oqs,
    author = "Chen, Cheng and Yan, Bing and Xie, Ju-Jun",
    title = "{Cross Sections and the Electromagnetic Form Factors within the Extended Vector Meson Dominance Model}",
    eprint = "2312.16753",
    archivePrefix = "arXiv",
    primaryClass = "hep-ph",
    doi = "10.1088/0256-307X/41/2/021302",
    journal = "Chin. Phys. Lett.",
    volume = "41",
    number = "2",
    pages = "021302",
    year = "2024"
}

@article{Xu:1992vc,
    author = "Xu, Q. P. and Kamal, A. N.",
    title = "{Cabibbo favored nonleptonic decays of charmed baryons}",
    reportNumber = "ALBERTA-THY-8-92",
    doi = "10.1103/PhysRevD.46.270",
    journal = "Phys. Rev. D",
    volume = "46",
    pages = "270--278",
    year = "1992"
}

@article{BESIII:2023wrw,
    author = "Ablikim, M. and others",
    collaboration = "BESIII Collaboration",
    title = "{First Measurement of the Decay Asymmetry in the Pure $W$-Boson-Exchange Decay $\Lambda_{c}^{+}\to\Xi^{0}K^{+}$}",
    eprint = "2309.02774",
    archivePrefix = "arXiv",
    primaryClass = "hep-ex",
    doi = "10.1103/PhysRevLett.132.031801",
    journal = "Phys. Rev. Lett.",
    volume = "132",
    number = "3",
    pages = "031801",
    year = "2024"
}

@article{Wan:2021ncg,
    author = "Wan, Junyao and Yang, Yongliang and Lu, Zhun",
    title = "{The electromagnetic form factors of $\Lambda_{c}$ hyperon in the vector meson dominance model}",
    eprint = "2102.03092",
    archivePrefix = "arXiv",
    primaryClass = "hep-ph",
    doi = "10.1140/epjp/s13360-021-01919-6",
    journal = "Eur. Phys. J. Plus",
    volume = "136",
    number = "9",
    pages = "949",
    year = "2021"
}

@article{Belle:2008xmh,
    author = "Pakhlova, G. and others",
    collaboration = "Belle Collaboration",
    title = "{Observation of a near-threshold enhancement in the $e^{+}e^{-}\to\Lambda^{+}_{c}\bar{\Lambda}^{-}_{c}$ cross section using initial-state radiation}",
    eprint = "0807.4458",
    archivePrefix = "arXiv",
    primaryClass = "hep-ex",
    doi = "10.1103/PhysRevLett.101.172001",
    journal = "Phys. Rev. Lett.",
    volume = "101",
    pages = "172001",
    year = "2008"
}

@article{Zhong:2024qqs,
    author = "Zhong, Huiling and Xu, Fanrong and Cheng, Hai-Yang",
    title = "{Analysis of hadronic weak decays of charmed baryons in the topological diagrammatic approach}",
    eprint = "2404.01350",
    archivePrefix = "arXiv",
    primaryClass = "hep-ph",
    doi = "10.1103/PhysRevD.109.114027",
    journal = "Phys. Rev. D",
    volume = "109",
    number = "11",
    pages = "114027",
    year = "2024"
}

@article{He:2024unv,
    author = "He, Xiao-Gang and Liu, Chia-Wei",
    title = "{Large $C\!P$ violation in charmed baryon decays}",
    eprint = "2404.19166",
    archivePrefix = "arXiv",
    primaryClass = "hep-ph",
    doi = "10.1016/j.scib.2025.05.045",
    journal = "Sci. Bull.",
    volume = "",
    number = "",
    pages = "",
    year = "2025",
}

@article{CLEO:1995qyd,
    author = "Bishai, M. and others",
    collaboration = "CLEO Collaboration",
    title = "{Measurement of the decay asymmetry parameters in $\Lambda_{c}^{+}\to\Lambda\pi^{+}$ and $\Lambda_{c}^{+}\to\Sigma^{+}\pi^{0}$}",
    eprint = "hep-ex/9502004",
    archivePrefix = "arXiv",
    reportNumber = "CLNS-95-1319, CLEO-95-1",
    doi = "10.1016/0370-2693(95)00280-X",
    journal = "Phys. Lett. B",
    volume = "350",
    pages = "256--262",
    year = "1995"
}

@article{Ivanov:1997ra,
    author = "Ivanov, Mikhail A. and Korner, J. G. and Lyubovitskij, Valery E. and Rusetsky, A. G.",
    title = "{Exclusive nonleptonic decays of bottom and charm baryons in a relativistic three quark model: Evaluation of nonfactorizing diagrams}",
    eprint = "hep-ph/9709372",
    archivePrefix = "arXiv",
    reportNumber = "MZ-TH-97-15",
    doi = "10.1103/PhysRevD.57.5632",
    journal = "Phys. Rev. D",
    volume = "57",
    pages = "5632--5652",
    year = "1998"
}

@article{Geng:2019xbo,
    author = "Geng, C. Q. and Liu, Chia-Wei and Tsai, Tien-Hsueh",
    title = "{Asymmetries of anti-triplet charmed baryon decays}",
    eprint = "1902.06189",
    archivePrefix = "arXiv",
    primaryClass = "hep-ph",
    doi = "10.1016/j.physletb.2019.05.024",
    journal = "Phys. Lett. B",
    volume = "794",
    pages = "19--28",
    year = "2019"
}

@article{Cheng:1993gf,
    author = "Cheng, Hai-Yang and Tseng, B.",
    title = "{Cabibbo allowed nonleptonic weak decays of charmed baryons}",
    eprint = "hep-ph/9304286",
    archivePrefix = "arXiv",
    reportNumber = "IP-ASTP-10-93, ITP-SB-93-20",
    doi = "10.1103/PhysRevD.48.4188",
    journal = "Phys. Rev. D",
    volume = "48",
    pages = "4188--4202",
    year = "1993"
}

@article{Cheng:1991sn,
    author = "Cheng, Hai-Yang and Tseng, B.",
    title = "{Nonleptonic weak decays of charmed baryons}",
    reportNumber = "IP-ASTP-17-91",
    doi = "10.1103/PhysRevD.46.1042",
    journal = "Phys. Rev. D",
    volume = "46",
    pages = "1042",
    year = "1992",
    note = "[Erratum: Phys.Rev.D 55, 1697 (1997)]"
}

@article{Korner:1992wi,
    author = "Korner, J. G. and Kramer, M.",
    title = "{Exclusive nonleptonic charm baryon decays}",
    reportNumber = "DESY-92-049, MZ-TH-91-07",
    doi = "10.1007/BF01561305",
    journal = "Z. Phys. C",
    volume = "55",
    pages = "659--670",
    year = "1992"
}

@article{Zenczykowski:1993hw,
    author = "Zenczykowski, P.",
    title = "{Quark and pole models of nonleptonic decays of charmed baryons}",
    eprint = "hep-ph/9309265",
    archivePrefix = "arXiv",
    reportNumber = "INP-1643-PH",
    doi = "10.1103/PhysRevD.50.402",
    journal = "Phys. Rev. D",
    volume = "50",
    pages = "402--411",
    year = "1994"
}

@article{Zenczykowski:1993jm,
    author = "Zenczykowski, Piotr",
    title = "{Nonleptonic charmed baryon decays: Symmetry properties of parity violating amplitudes}",
    reportNumber = "INP-1655-PH",
    doi = "10.1103/PhysRevD.50.5787",
    journal = "Phys. Rev. D",
    volume = "50",
    pages = "5787--5792",
    year = "1994"
}

@MISC{Datta:1995mn,
    author = "Datta, Alakabha",
    title = "{Nonleptonic two-body decays of charmed and $\Lambda_{b}$ baryons}",
    howpublished = "e-print arXiv:hep-ph/9504428", 
    eprint = "hep-ph/9504428",
    archivePrefix = "arXiv",
    reportNumber = "UH-511-824-95",
    month = "4",
    year = "1995"
}

@article{Sharma:1998rd,
    author = "Sharma, K. K. and Verma, R. C.",
    title = "{A Study of weak mesonic decays of $\Lambda_{c}$ and $\Xi_{c}$ baryons on the basis of HQET results}",
    eprint = "hep-ph/9803302",
    archivePrefix = "arXiv",
    doi = "10.1007/s100529801008",
    journal = "Eur. Phys. J. C",
    volume = "7",
    pages = "217--224",
    year = "1999"
}

@article{CLEO:1990unw,
    author = "Avery, P. and others",
    collaboration = "CLEO Collaboration",
    title = "{Measurement of the $\Lambda_{c}^{+}$ decay asymmetry parameter}",
    reportNumber = "CLNS-90-1013, CLEO-90-10",
    doi = "10.1103/PhysRevLett.65.2842",
    journal = "Phys. Rev. Lett.",
    volume = "65",
    pages = "2842--2845",
    year = "1990"
}

@article{ARGUS:1991yzs,
    author = "Albrecht, H. and others",
    collaboration = "ARGUS Collaboration",
    title = "{A Measurement of asymmetry in the decay $\Lambda_{c}^{+}\to\Lambda\pi^{+}$}",
    reportNumber = "DESY-91-091",
    doi = "10.1016/0370-2693(92)90529-D",
    journal = "Phys. Lett. B",
    volume = "274",
    pages = "239--245",
    year = "1992"
}

@article{FOCUS:2005vxq,
    author = "Link, J. M. and others",
    collaboration = "FOCUS Collaboration",
    title = "{Study of the decay asymmetry parameter and $C\!P$ violation parameter in the $\Lambda_{c}^{+}\to\Lambda\pi^{+}$ decay}",
    eprint = "hep-ex/0509042",
    archivePrefix = "arXiv",
    reportNumber = "FERMILAB-PUB-05-424-E",
    doi = "10.1016/j.physletb.2006.01.017",
    journal = "Phys. Lett. B",
    volume = "634",
    pages = "165--172",
    year = "2006"
}

@article{LHCb:2024tnq,
    author = "Aaij, Roel and others",
    collaboration = "LHCb Collaboration",
    title = "{Measurement of $\it{\Lambda^\text{0}_b}$, $\it{\Lambda^\text{+}_c}$, and $\it{\Lambda}$ Decay Parameters Using $\it{\Lambda^\text{0}_b} \to \it{\Lambda^\text{+}_c} h^-$ Decays}",
    eprint = "2409.02759",
    archivePrefix = "arXiv",
    primaryClass = "hep-ex",
    reportNumber = "LHCb-PAPER-2024-017, CERN-EP-2024-200",
    doi = "10.1103/PhysRevLett.133.261804",
    journal = "Phys. Rev. Lett.",
    volume = "133",
    number = "26",
    pages = "261804",
    year = "2024"
}

@article{Pacetti:2014jai,
    author = "Pacetti, S. and Baldini Ferroli, R. and Tomasi-Gustafsson, E.",
    title = "{Proton electromagnetic form factors: Basic notions, present achievements and future perspectives}",
    doi = "10.1016/j.physrep.2014.09.005",
    journal = "Phys. Rept.",
    volume = "550-551",
    pages = "1--103",
    year = "2015"
}

@article{Huang:2021xte,
    author = "Huang, Guangshun and Ferroli, Rinaldo Baldini",
    collaboration = "BESIII Collaboration",
    title = "{Probing the internal structure of baryons}",
    eprint = "2111.08425",
    archivePrefix = "arXiv",
    primaryClass = "hep-ex",
    doi = "10.1093/nsr/nwab187",
    journal = "Natl. Sci. Rev.",
    volume = "8",
    number = "11",
    pages = "nwab187",
    year = "2021"
}

@article{BESIII:2023cvk,
    author = "Ablikim, Medina and others",
    collaboration = "BESIII Collaboration",
    title = "{Extracting the femtometer structure of strange baryons using the vacuum polarization effect}",
    eprint = "2309.04139",
    archivePrefix = "arXiv",
    primaryClass = "hep-ex",
    doi = "10.1038/s41467-024-51802-y",
    journal = "Nature Commun.",
    volume = "15",
    number = "1",
    pages = "8812",
    year = "2024"
}

@article{Cheng:2024lsn,
    author = "Cheng, Hai-Yang and Xu, Fanrong and Zhong, Huiling",
    title = "{Hadronic weak decays of charmed baryons in the topological diagrammatic approach: An update}",
    eprint = "2410.04675",
    archivePrefix = "arXiv",
    primaryClass = "hep-ph",
    doi = "10.1103/PhysRevD.111.034011",
    journal = "Phys. Rev. D",
    volume = "111",
    number = "3",
    pages = "034011",
    year = "2025"
}

@article{Wang:2024wrm,
    author = {Hong-Jian Wang and Pei-Rong Li and Xiao-Rui Lyu and Jusak Tandean and Hai-Bo Li},
    title = {Remarks on strong phase shifts in weak nonleptonic baryon decays},
    eprint = "2412.02170",
    archivePrefix = "arXiv",
    primaryClass = "hep-ph",
    doi = "10.1016/j.scib.2025.02.030",
    journal = "Sci. Bull.",
    volume = "70",
    pages = "1189--1191",
    year = "2025"
}

@article{KTeV:1999kad,
    author = "Alavi-Harati, A. and others",
    collaboration = "KTeV Collaboration",
    title = "{Observation of Direct $C\!P$ Violation in $K_{S,L} \to \pi \pi$ Decays}",
    eprint = "hep-ex/9905060",
    archivePrefix = "arXiv",
    reportNumber = "EFI-99-25, FERMILAB-PUB-99-150-E",
    doi = "10.1103/PhysRevLett.83.22",
    journal = "Phys. Rev. Lett.",
    volume = "83",
    pages = "22--27",
    year = "1999"
}

@article{LHCb:2024yzj,
    author = "Aaij, Roel and others",
    collaboration = "LHCb Collaboration",
    title = "{Study of $\Lambda_b^0$ and $\Xi_b^0$ Decays to $\Lambda h^+h'^-$ and Evidence for $C\!P$ Violation in $\Lambda_b^0\to\Lambda K^+K^-$ Decays}",
    eprint = "2411.15441",
    archivePrefix = "arXiv",
    primaryClass = "hep-ex",
    reportNumber = "LHCb-PAPER-2024-043, CERN-EP-2024-281",
    doi = "10.1103/PhysRevLett.134.101802",
    journal = "Phys. Rev. Lett.",
    volume = "134",
    number = "10",
    pages = "101802",
    year = "2025"
}

@article{LHCb:2016yco,
    author = "Aaij, Roel and others",
    collaboration = "LHCb Collaboration",
    title = "{Measurement of matter-antimatter differences in beauty baryon decays}",
    eprint = "1609.05216",
    archivePrefix = "arXiv",
    primaryClass = "hep-ex",
    reportNumber = "CERN-EP-2016-212, LHCB-PAPER-2016-030",
    doi = "10.1038/nphys4021",
    journal = "Nature Phys.",
    volume = "13",
    pages = "391--396",
    year = "2017"
}

@article{LHCb:2025ray,
    author = "Aaij, Roel and others",
    collaboration = "LHCb Collaboration",
    title = "{Observation of charge-parity symmetry breaking in baryon decays}",
    eprint = "2503.16954",
    archivePrefix = "arXiv",
    primaryClass = "hep-ex",
    reportNumber = "LHCb-PAPER-2024-054, CERN-EP-2025-031",
    doi = "10.1038/s41586-025-09119-3",
    journal = "Nature",
    volume = "643",
    pages = "1223--1228",
    year = "2025"
}

@article{Wang:2025yjs,
    author = "Wang, Hong-Jian and Wang, Cheng and Sun, Hao and Li, Pei-Rong and Lyu, Xiao-Rui and Ping, Rong-Gang",
    title = "{Prospects for $P$ and $C\!P$ violation in $\Lambda^+_c$ decays with a polarized beam at the Super Tau-Charm Facility}",
    eprint = "2508.12217",
    archivePrefix = "arXiv",
    primaryClass = "hep-ph",
    doi = "10.1103/6pdw-trsj",
    journal = "Phys. Rev. D",
    volume = "112",
    number = "11",
    pages = "112002",
    year = "2025"
}

@article{Li:2025nzx,
    author = "Li, Pei-Rong and Lyu, Xiao-Rui and Zheng, Yangheng",
    title = "{Experimental overview on the charmed baryon decays}",
    eprint = "2509.19141",
    archivePrefix = "arXiv",
    primaryClass = "hep-ex",
    doi = "10.1088/1674-1137/ae1187",
    journal = "Chin. Phys. C",
    volume = "50",
    number = "2",
    pages = "022002",
    year = "2026"
}

@article{Abrams:1979iu,
    author = "Abrams, G. S. and others",
    title = "{Observation of Charmed Baryon Production in $e^+e^-$ Annihilation}",
    reportNumber = "SLAC-PUB-2406, LBL-9855",
    doi = "10.1103/PhysRevLett.44.10",
    journal = "Phys. Rev. Lett.",
    volume = "44",
    pages = "10",
    year = "1980"
}

@article{Schonning:2023mge,
    author = {Sch{\"o}nning, Karin and Batozskaya, Varvara and Adlarson, Patrik and Zhou, Xiaorong},
    title = "{Production and decay of polarized hyperon-antihyperon pairs}",
    eprint = "2302.13071",
    archivePrefix = "arXiv",
    primaryClass = "hep-ph",
    doi = "10.1088/1674-1137/acc790",
    journal = "Chin. Phys. C",
    volume = "47",
    number = "5",
    pages = "052002",
    year = "2023"
}

@article{Langenbruch:2019nwe,
    author = "Langenbruch, Christoph",
    title = "{Parameter uncertainties in weighted unbinned maximum likelihood fits}",
    eprint = "1911.01303",
    archivePrefix = "arXiv",
    primaryClass = "physics.data-an",
    doi = "10.1140/epjc/s10052-022-10254-8",
    journal = "Eur. Phys. J. C",
    volume = "82",
    number = "5",
    pages = "393",
    year = "2022"
}

@article{BaBar:2004gyj,
    author = "Aubert, Bernard and others",
    collaboration = "BaBar Collaboration",
    title = "{Observation of direct CP violation in $B^0 \to K^+ \pi^-$ decays}",
    eprint = "hep-ex/0407057",
    archivePrefix = "arXiv",
    reportNumber = "SLAC-PUB-10582, BABAR-PUB-04-037",
    doi = "10.1103/PhysRevLett.93.131801",
    journal = "Phys. Rev. Lett.",
    volume = "93",
    pages = "131801",
    year = "2004"
}

@article{Li:2021lvs,
    author = "Li, Zhong-Yi and Dai, An-Xin and Xie, Ju-Jun",
    title = "{Electromagnetic Form Factors of {\ensuremath{\Lambda}} Hyperon in the Vector Meson Dominance Model and a Possible Explanation of the Near-Threshold Enhancement of the Reaction}",
    eprint = "2107.10499",
    archivePrefix = "arXiv",
    primaryClass = "hep-ph",
    doi = "10.1088/0256-307X/39/1/011201",
    journal = "Chin. Phys. Lett.",
    volume = "39",
    number = "1",
    pages = "011201",
    year = "2022"
}

@article{Chen:2024luh,
    author = "Chen, Cheng and Yan, Bing and Xie, Ju-Jun",
    title = "{The electromagnetic form factors and spin polarization of $\Lambda_c^+$ in the process $e^+e^-\to\Lambda_c^+\bar{\Lambda}_c^-$}",
    eprint = "2407.19445",
    archivePrefix = "arXiv",
    primaryClass = "hep-ph",
    doi = "10.1088/1674-1137/ad9259",
    journal = "Chin. Phys. C",
    volume = "49",
    number = "2",
    pages = "023102",
    year = "2025"
}

@article{Yang:2019mzq,
    author = "Yang, Yongliang and Chen, Dian-Yong and Lu, Zhun",
    title = "{Electromagnetic form factors of $\Lambda$ hyperon in the vector meson dominance model}",
    eprint = "1902.01242",
    archivePrefix = "arXiv",
    primaryClass = "hep-ph",
    doi = "10.1103/PhysRevD.100.073007",
    journal = "Phys. Rev. D",
    volume = "100",
    number = "7",
    pages = "073007",
    year = "2019"
}

@article{BESIII:2022udq,
    author = "Ablikim, Medina and others",
    collaboration = "BESIII Collaboration",
    title = "{Partial wave analysis of the charmed baryon hadronic decay $ {\Lambda}_c^{+} ${\textrightarrow} {\ensuremath{\Lambda}}{\ensuremath{\pi}}$^{+}${\ensuremath{\pi}}$^{0}$}",
    eprint = "2209.08464",
    archivePrefix = "arXiv",
    primaryClass = "hep-ex",
    doi = "10.1007/JHEP12(2022)033",
    journal = "JHEP",
    volume = "12",
    pages = "033",
    year = "2022"
}

@article{Jia:2024pyb,
    author = "Jia, Cai-Ping and Jiang, Hua-Yu and Wang, Jian-Peng and Yu, Fu-Sheng",
    title = "{Final-state rescattering mechanism of charmed baryon decays}",
    eprint = "2408.14959",
    archivePrefix = "arXiv",
    primaryClass = "hep-ph",
    doi = "10.1007/JHEP11(2024)072",
    journal = "JHEP",
    volume = "11",
    pages = "072",
    year = "2024"
}

@article{NA48:1999szy,
    author = "Fanti, V. and others",
    collaboration = "NA48 Collaboration",
    title = "{A New measurement of direct CP violation in two pion decays of the neutral kaon}",
    eprint = "hep-ex/9909022",
    archivePrefix = "arXiv",
    reportNumber = "CERN-EP-99-114",
    doi = "10.1016/S0370-2693(99)01030-8",
    journal = "Phys. Lett. B",
    volume = "465",
    pages = "335--348",
    year = "1999"
}

@article{Belle:2004nch,
    author = "Chao, Y. and others",
    collaboration = "Belle Collaboration",
    title = "{Evidence for direct CP violation in $B^0\to K^+\pi^-$ decays}",
    eprint = "hep-ex/0408100",
    archivePrefix = "arXiv",
    doi = "10.1103/PhysRevLett.93.191802",
    journal = "Phys. Rev. Lett.",
    volume = "93",
    pages = "191802",
    year = "2004"
}

@article{Yang:2025orn,
    author = "Yang, Chang and He, Xiao-Gang and Liu, Chia-Wei",
    title = "{Charmed baryon decays: SU(3)$_{F}$ breaking and CP violation}",
    eprint = "2506.19005",
    archivePrefix = "arXiv",
    primaryClass = "hep-ph",
    doi = "10.1007/JHEP09(2025)193",
    journal = "JHEP",
    volume = "09",
    pages = "193",
    year = "2025"
}

@article{BESIII:2019nep,
    author = "Ablikim, Medina and others",
    collaboration = "BESIII Collaboration",
    title = "{Complete Measurement of the $\Lambda$ Electromagnetic Form Factors}",
    eprint = "1903.09421",
    archivePrefix = "arXiv",
    primaryClass = "hep-ex",
    doi = "10.1103/PhysRevLett.123.122003",
    journal = "Phys. Rev. Lett.",
    volume = "123",
    number = "12",
    pages = "122003",
    year = "2019"
}

@article{BESIII:2025kyg,
    author = "Ablikim, Medina and others",
    collaboration = "BESIII Collaboration",
    title = "{Unraveling the structure of $\Lambda$ hyperons with polarized $\Lambda\bar{\Lambda}$ pairs}",
    eprint = "2506.08072",
    archivePrefix = "arXiv",
    primaryClass = "hep-ex",
    doi = "10.1103/9f1h-fl4k",
    journal = "Phys. Rev. Lett.",
    volume = "135",
    pages = "191902",
    year = "2025"
}

@article{Salone:2022lpt,
    author = "Salone, Nora and Adlarson, Patrik and Batozskaya, Varvara and Kupsc, Andrzej and Leupold, Stefan and Tandean, Jusak",
    title = "{Study of $C\!P$ violation in hyperon decays at super-charm-tau factories with a polarized electron beam}",
    eprint = "2203.03035",
    archivePrefix = "arXiv",
    primaryClass = "hep-ph",
    doi = "10.1103/PhysRevD.105.116022",
    journal = "Phys. Rev. D",
    volume = "105",
    number = "11",
    pages = "116022",
    year = "2022"
}

@article{Mangoni:2021qmd,
    author = "Mangoni, Alessio and Pacetti, Simone and Tomasi-Gustafsson, Egle",
    title = "{The first exploration of the physical Riemann surfaces of the ratio $G_E^\Lambda/G_M^\Lambda$}",
    eprint = "2109.03759",
    archivePrefix = "arXiv",
    primaryClass = "hep-ph",
    doi = "10.1103/PhysRevD.104.116016",
    journal = "Phys. Rev. D",
    volume = "104",
    number = "11",
    pages = "116016",
    year = "2021"
}

@article{Punjabi:2015bba,
    author = "Punjabi, V. and Perdrisat, C. F. and Jones, M. K. and Brash, E. J. and Carlson, C. E.",
    title = "{The Structure of the Nucleon: Elastic Electromagnetic Form Factors}",
    eprint = "1503.01452",
    archivePrefix = "arXiv",
    primaryClass = "nucl-ex",
    reportNumber = "JLAB-PHY-15-2019",
    doi = "10.1140/epja/i2015-15079-x",
    journal = "Eur. Phys. J. A",
    volume = "51",
    pages = "79",
    year = "2015"
}

@article{Atac:2021wqj,
    author = "Atac, H. and Constantinou, M. and Meziani, Z. -E and Paolone, M. and Sparveris, N.",
    title = "{Measurement of the neutron charge radius and the role of its constituents}",
    eprint = "2103.10840",
    archivePrefix = "arXiv",
    primaryClass = "nucl-ex",
    doi = "10.1038/s41467-021-22028-z",
    journal = "Nature Commun.",
    volume = "12",
    number = "1",
    pages = "1759",
    year = "2021"
}

@MISC{BESIII:2026mye,
    author = "Ablikim, Medina and others",
    collaboration = "BESIII Collaboration",
    title = "{Precision Studies and Searches for $C\!P$ Asymmetries in the Inclusive Decay $\Lambda^+_c\to \Lambda X$}",
    howpublished = "e-print arXiv:2602.24089",
    eprint = "2602.24089",
    archivePrefix = "arXiv",
    primaryClass = "hep-ex",
    month = "2",
    year = "2026"
}
